\documentclass[a4paper,12pt]{article}
\usepackage[
left=1.1in,
right=1.1in,
top=1.in,
bottom=1.in,
						]{geometry} 

\usepackage{jheppub} 

\usepackage[T1]{fontenc} 

\usepackage{amsfonts}
\usepackage{amssymb}
\usepackage{dsfont}		

\input epsf

\usepackage[shortlabels]{enumitem}

\usepackage{braket}

\makeatletter
\newenvironment{sqcases}{%
  \matrix@check\sqcases\env@sqcases
}{%
  \endarray\right.%
}
\def\env@sqcases{%
  \let\@ifnextchar\new@ifnextchar
  \left\lbrack
  \def\arraystretch{1.2}%
  \array{@{}l@{\quad}l@{}}%
}
\makeatother

\usepackage[titletoc]{appendix}

\usepackage{enumitem}

\hypersetup{
    colorlinks=true,
    anchorcolor=red,
    urlcolor=magenta,
    citecolor=red,
    linkcolor=cyan}
        
\usepackage{simplewick}    

\usepackage{empheq}

\usepackage{mathrsfs}

\usepackage{stmaryrd}

\usepackage{subfigure}

\def\a{\alpha}
\def\b{\beta}
\def\s{\sigma}

\def\la{\lambda}
\def\La{\Lambda}

\def\ga{\gamma}

\def\vare{\varepsilon}
\def\e{\epsilon}

\def\rm{\mathrm}
\def\cal{\mathcal}
\def\scr{\mathscr}

\def\pa{\partial}

\def\be{\begin{equation}}
\def\ee{\end{equation}}
\def\br{\begin{eqnarray}}

\def\er{\end{eqnarray}}
\def\bsub{\begin{subequations}}
\def\esub{\end{subequations}}

\def\NS{\mathrm{NS}}
\def\R{\mathrm{R}}

\def\lor{\, \colon \!}
\def\ror{\! \colon \,}

\def\Oint{O^{(\rm{int})}_{[2]}}

\def\Oincov{O^{(\rm{int})}}

\def\inter{\rm{int}}
\def\free{\rm{free}}

\title{\boldmath Renormalization of Twisted Ramond Fields in D1-D5 SCFT$_2$ }

\author[a,1]{A.~A.~Lima,\note{Corresponding author.}}
\author[a]{G.~M.~Sotkov}
\author[b]{M.~Stanishkov}

\affiliation[a]{Department of Physics, Federal University of Esp\'irito Santo,\\ 29075-900, Vit\'oria, Brazil}
\affiliation[b]{Institute for Nuclear Research and Nuclear Energy,\\ Bulgarian Academy of Sciences, 1784 Sofia, Bulgaria}

\emailAdd{andrealves.fis@gmail.com}
\emailAdd{gsotkov@gmail.com}
\emailAdd{marian@inrne.bas.bg}

\abstract{
We explore the $n$-twisted Ramond sector of the deformed two-dimensional $\cal N = (4, 4)$ superconformal $(T^4)^N /S_N$ orbifold theory, describing bound states of D1-D5 brane system in type IIB superstring. We derive the large-$N$ limit of the four-point function of two R-charged twisted Ramond fields and two marginal deformation operators at the free orbifold point. Specific short-distance limits of this function provide several structure constants, the OPE fusion rules and the conformal dimensions of a few non-BPS operators. The second order correction (in the deformation parameter) to the two-point function of the Ramond fields, defined as double integrals over this four-point function, turns out to be UV-divergent, requiring an appropriate renormalization of the fields. We calculate the corrections to the conformal dimensions of the twisted Ramond ground states at the large-$N$ limit. The same integral yields the first-order deviation from zero of the structure constant of the three-point function of two Ramond fields and one deformation operator. Similar results concerning the correction to the two-point function of bare twist operators and their renormalization are also obtained. }


\begin{document} 
\maketitle
\flushbottom

\section{Introduction} \label{sec:Introduction}%

%
%


In the near-horizon decoupling limit \cite{Maldacena:1997re} of the type-IIB supergravity description of bound states of $N_1$ D1- and $N_5$ D5-branes, the asymptotic geometry becomes $AdS_3 \times S^3 \times T^4$, with large Ramond-Ramond charges \cite{Maldacena:1998bw,Seiberg:1999xz}, from which one can reconstruct its holographic dual   SCFT$_2$; see  \cite{David:2002wn,Mathur:2005zp,Skenderis:2008qn} for reviews.
There is strong indication that this D1-D5 SCFT$_2$ flows in the infrared to a free field theory whose sigma model is $(T^4)^N/S_N$, an orbifold of $T^4$  by the symmetric group $S_N$, with  $N = N_1 N_5$, 
while the supergravity description is obtained by moving in moduli space with a deformation away from this `free orbifold point'. 
Gravitational solutions, which include the Strominger-Vafa black hole \cite{Strominger:1996sh} and fuzzball geometries \cite{Lunin:2001jy,Mathur:2005zp,Kanitscheider:2007wq,Kanitscheider:2006zf,Skenderis:2008qn,Mathur:2018tib}, 
are all dual to states in the Ramond sector of the SCFT$_2$, and one can find correspondences between the geometries and Ramond states in many cases where the latter are BPS-protected from renormalization.
Extensive research has achieved considerable progress in the understanding of the free orbifold and its deformation, as well as in the construction of `superstratum' geometries corresponding to the microscopic picture 
\cite{Lunin:2000yv,Lunin:2001pw,Balasubramanian:2005qu,Avery:2009tu,Pakman:2009ab,Pakman:2009mi,Burrington:2012yq,Bena:2013dka,Carson:2014ena,Carson:2015ohj,Fitzpatrick:2016ive,Burrington:2017jhh,Galliani:2017jlg,Bombini:2017sge,Tormo:2018fnt, Eberhardt:2018ouy, Bena:2019azk,Dei:2019osr,Eberhardt:2019ywk,Giusto:2018ovt,Martinec:2019wzw,Hampton:2019csz,Warner:2019jll,Dei:2019iym,Galliani:2016cai,Bombini:2019vnc,Giusto:2020mup}.
Nevertheless,  the description of the dynamics of the deformed SCFT$_2$ is still not fully understood. One of the open problems concerns  the selection rules separating protected states from ``lifted'' ones, whose conformal data flow in the deformed theory after renormalization \cite{Guo:2019pzk,Keller:2019suk,Keller:2019yrr,Guo:2019ady,Belin:2019rba,Guo:2020gxm}.

The  present  paper investigates  the effects on the conformal properties  of   twisted ground states in the $\cal N = (4,4)$ orbifold SCFT$_2$ when the theory is deformed by a marginal scalar modulus operator $\lambda \Oint$  \cite{Avery:2010er,Avery:2010hs,Carson:2014ena,Carson:2015ohj,Carson:2016uwf}.  
The first-order correction, in powers of $\lambda$, of the two-point function of a ground state is known to vanish. 
Our main result is an explicit derivation of the finite part of the second-order correction to two-point functions of $n$-twisted primary operators $\scr O_{[n]}$, by eliminating the UV divergences with an appropriate renormalization of the fields.  As a consequence, the scaling dimension $\Delta^{\scr O}_n(0)$ in the free orbifold point flows with $\la$ according to 
\be
\Delta_n^{\scr O}(\la)= \Delta^{\scr O}_n(0) + \frac{\pi}{2} \la^2 | J_{\scr O}(n)|, \label{anomdimintro}
\ee
where  $J_{\scr O}(n)$ is a regularized integral defined in Sect.\ref{SectRenormOfConf} below. 
%
%
%
%
We will be mostly interested in two specific operators: the bare twist field $\s_{[n]}$, and, more importantly, the twisted R-charged Ramond fields $R^\pm_{[n]}$, with bare (holomorphic) conformal weight $h = \frac{n}{4}$. 
In both cases, our main result --- the lifting of the conformal dimensions --- holds for $n < N$.
Ramond fields with $n=N$ and $n=2$, i.e.~with maximal and minimal twist,  are protected at leading order in the large-$N$ approximation. Bare twist fields with $n = N$ are also protected at leading order.
For the particular case of  twisted Ramond fields, this has been recently reported in our short letter \cite{Lima:2020boh}.

The main ingredient  in the calculation of second-order corrections to the two-point functions is finding an explicit analytic expression for the four-point function
\be
\big\langle \scr O_{[n]}^\dagger(z_1,\bar z_1) \Oint(z_2, \bar z_2) \Oint (z_3 , \bar z_3) \scr O_{[n]}(z_4,\bar z_4) \big\rangle 
\label{4ptfuncintro}
\ee
%
%
%
for the fields $\scr O_{[n]}$ we are interested in.
We will present a detailed derivation of the large-$N$ approximation of the corresponding functions,  by applying covering surface techniques \cite{Lunin:2000yv,Lunin:2001pw} combined with the `stress-tensor method' \cite{Dixon:1985jw}. Our computation of the leading term in the $1/N$ expansion of the connected part of (\ref{4ptfuncintro}), takes into account  only the terms contributing to genus zero  surfaces, i.e.~we use  the well known map \cite{Arutyunov:1997gi,Arutyunov:1997gt} between the ``base'' branched sphere to its genus-zero covering surface \cite{Lunin:2000yv}.
The alliance of the covering surface with the stress-tensor method emphasizes some interesting mathematical properties of the correlation functions, and their relation to Hurwitz theory \cite{Pakman:2009ab,Pakman:2009zz,Pakman:2009mi}.

Corrections to the anomalous dimensions at second-order follow from the integral of (\ref{4ptfuncintro}) over the positions of the interaction operators.
The analogous integral in the case where there are NS chiral fields at $z_1$ and $z_4$ has been computed in \cite{Pakman:2009mi}, and shown to vanish, as expected for protected operators which should not renormalize.
For the twisted Ramond fields and the twist operators, however, the integrand has a more complicated structure, with one more branch cut,  and without appropriate regularization the integrals are divergent.  In order to define and evaluate their finite parts, we have elaborated a regularization procedure and a specific renormalization scheme for the fields in the deformed SCFT$_2$.
Our starting point is the observation that the integrals  we are interested in can be put in a form  studied by Dotsenko and Fateev \cite{Dotsenko:1984nm,Dotsenko:1984ad,dotsenko1988lectures} in a different context, as integral representation of the conformal blocks  of  primary fields (curiously; not of their integrals) in the $c<1$ series%
		\footnote{See Refs.\cite{Mussardo:1987eq,Mussardo:1987ab,Mussardo:1988av} for the extension  of the Dotsenko-Fateev integral representation to the Ramond and twisted sectors of  the $\cal N=1$ and $\cal N=2$ supersymmetric minimal models.}
of minimal  CFT$_2$ models. 
While, in the one hand, they can be formally written as specific contour integrals in the complex plane --- with the contours ensuring a series of algebraic properties --- on the other hand these integrals can be represented by four `canonical functions' which  are analytic in their parameters, even in cases where the integral itself diverges. Thus, by analytic continuation, the canonical functions give a regularized result for the desired integrals of the four-point functions (\ref{4ptfuncintro}).  
When applied to the parameters of NS chirals, this procedure gives a vanishing result, as expected; but when applied to the Ramond and twisted fields, we find finite, non-vanishing corrections to the conformal dimensions.
The analytic expressions for the renormalized conformal dimensions of $R^\pm_{[n]}$ and $\s_{[n]}$ is one of the most important results of this paper.
As a byproduct of the computation of the integrals, we can also present the first-order correction to the structure constants
$\langle R_{[n]}^-(\infty) \Oint (1) R_{[n]}^+(0) \rangle$,
and
$\langle \s_{[n]}(\infty) \Oint (1) \s_{[n]}(0) \rangle$,
 which do not vanish in the deformed SCFT$_2$.  It is worthwhile to mention the recent  use of similar methods to the renormalization of  certain \emph{composite} Ramond fields, for example $R_{[n]}^+(z)R^-_{[m]}(z)$ \cite{Lima:2020nnx}. In the composite case, an important consequence of the renormalization procedure is  the existence of a condition, namely $m+n=N$, selecting a class of protected (non-renormalized) states. The remaining states, with $n+m<N$, are lifted: their renormalized conformal dimensions flow with $\la$, and are given by the sum of the second-order corrections (\ref{anomdimintro}) for each one of the constituents, i.e.~$\Delta_n^R(\la)  +\Delta_m^R(\la)$.
What distinguishes the protected Ramond fields from the lifted ones is that the former have conformal weight $h = \frac1{24} c_{orb} = \frac14 N$, i.e.~they are Ramond ground states of the full orbifold theory; in contrast, the fields with $h = \frac14 n$, $n < N$ are Ramond ground states only of the $n$-twisted sector, or of the `$n$-wound component string' in the familiar description of the D1-D5 SCFT in terms of effective stings.

Another contribution of the present paper is the analysis of short-distance limits  of the four-point function (\ref{4ptfuncintro}).
In the limits where operators coincide, $u \to 0,1,\infty$, we are able to derive  several structure constants, the OPE fusion rules  and the conformal dimensions of some non-BPS operators.
These OPE data add to the description of  the Ramond and twisted sectors of the free-orbifold point. Our results for the non-BPS fields are consistent  with what is known about the chiral NS and twisted sectors \cite{Pakman:2009mi,Burrington:2017jhh}. They are also in agreement with the recently conjectured universality of OPEs of certain chiral fields and the deformation operator in the large-$N$ limit \cite{Burrington:2018upk,deBeer:2019ioe}, and represent an extension of these results for all other sectors of the free orbifold theory.
In particular, we find that the OPE algebra of the deformation operator and the Ramond fields includes a set of R-charged twisted non-BPS operators $Y^\pm_{m}$, appearing in the OPEs $\Oint(z,\bar z) R^\pm_n(0)$.  
Similarly, the algebra of $\Oint$ and $\s_n$ includes new twisted operators $\cal Y_{m}$. We have calculated  the dimensions of these operators, as well as the values of structure constants such as $\langle R^\pm_n(\infty) \Oincov_2(1) Y^\pm_{n\pm1}(0) \rangle$.
Applying the fractional spectral flows of Ref.\cite{deBeer:2019ioe}  with $\xi=n/(n+1)$, we find that our results for the twisted Ramond fields' OPEs are in complete correspondence with those obtained from OPEs in the NS sector resulting in specific non-BPS  NS fields.

The structure of the paper is as follows.
In Sects.\ref{orbi} and \ref{SectRenormConfData}, we fix our notations by defining first the free orbifold SCFT$_2$, and then its deformation away from the free orbifold point; we also review some key features of conformal perturbation theory used later.
In Sect.\ref{SectFourPointFuncts},  we give a detailed calculation of the four-point functions involving Ramond and bare twist fields, necessary for the second-order correction of the two-point functions.
In Sect.\ref{SectOPElimits}, we investigate certain short-distance  limits of the four-point function in order to extract OPE fusion rules, conformal weights and structure constants of several operators in the free-orbifold point.
In Sect.\ref{SectRenormOfConf}, we return to conformal perturbation theory, with a detailed study of the regularization and the final computation of integrals resulting in the change of the conformal weights of $R^\pm_{[n]}$ and $\s_{[n]}$; we also explain how the renormalization scheme can be extended to a generic primary field $\scr O_{[n]}$; 
we also comment on the spectral flow between the fields $R^\pm_{[n]}$ and NS chiral operators in the free theory, and how it is ``broken'' after the deformation.
In Sect.\ref{SectConclusion} we present a compact summary of our results, together with a short discussion of a few open problems and the eventual consequences of  the continuous ($\lambda$-dependent)  conformal dimensions of the  renormalized twisted Ramond fields for their  geometric  bulk counterparts.
Some auxiliary topics are left for the appendices.

\section{D1-D5 SCFT$_2$ and $(T^4)^N/S_N$  orbifold} \label{orbi}

The `free orbifold point' of the D1-D5 system is the SCFT$_2$ with central charge 
 $(c_{orb}, \tilde c_{orb}) = (6N,6N)$, obtained by taking $N$ copies of the 
free $\cal N = (4,4)$ SCFT$_2$, identified under the symmetric group $S_N$, with target space $(T^4)^N/S_N$.
The $\cal N = (4,4)$ superconformal algebra of the `seed theory' has central charge $(c, \tilde c) = (6,6)$, R-symmetry group
$\rm{SU(2)}_L \times \rm{SU(2)}_R$, 
and
`internal' group $\rm{SO(4)_I} = \rm{SU(2)}_1 \times \rm{SU(2)}_2$ corresponding to the torus $T^4$ of target space. 
We work on the complex plane, with coordinates $z, \bar z$. 

The unitary representations of the holomorphic $\cal {N}= 4$ algebra are characterized by three  numbers $\{h, j^3,\frak j^3$\}, respectively the conformal weight and the semi-integer charges under the R-current $J^3(z)$ of SU(2)$_L$, and a current $\frak J^3(z)$ of the  global SU(2)$_1$. Similar numbers $\{ \tilde h, \tilde j^3, \tilde {\frak j}^3\}$ characterize the anti-holomorphic sector with SU(2)$_R$ and SU(2)$_2$ groups. 

The theory can be realized in terms of free bosons $X^{\dot A A}(z, \bar z)$ and free fermions $\psi^{\a \dot A}(z)$, $\tilde \psi^{\dot \a \dot A}(\bar z)$, whereas the stress-tensor, the R-current and the super-current are expressed as
\bsub\label{TJGasXpsi}
\begin{align}
T(z) &= \tfrac{1}{4} \e_{\dot A \dot B} \e_{AB} \pa X^{\dot A A} \pa X^{\dot B B} + \tfrac{1}{4} \e_{\dot A \dot B} \e_{\a\b} \psi^{\a \dot A} \pa \psi^{\b \dot B}  \label{TJGasXpsiT}
\\
J^a(z) &= \tfrac{1}{4} \e_{\dot A \dot B} \e_{\a\b} \psi^{\a \dot A} [\s^{*a}]^\b{}_\ga \psi^{\ga \dot B}
\\
G^{\a A}(z) &= \e_{\dot A \dot B} \psi^{\a \dot A} \pa X^{\dot B A}
\end{align}\esub
with similar expressions for the anti-holomorphic sector.
Conventions for SU(2) indices are given in Appendix \ref{AppConventions}.
The complex bosons and the complex fermions obey reality conditions (\ref{RealiCondXpsi}), 
and can be written in terms of real bosons and fermions $X_i(z,\bar z)$, $\psi_i(z)$ and $\tilde \psi_i(\bar z)$, $i= 1,2,3,4$; see (\ref{DoublFermiDef}).
The fermions can be described in terms of chiral scalar bosons $\phi_r(z)$ and $\tilde \phi_r(\bar z)$, with $r=1,2$. In the holomorphic sector,
\be
	\begin{bmatrix}
	\psi^{+ \dot 1} (z) \\ \psi^{- \dot 1} (z) 
	\end{bmatrix}
	=
	\begin{bmatrix}
	 e^{- i \phi_2(z)} \\ e^{- i \phi_1(z)}  
	\end{bmatrix} \ ,
\qquad
	\begin{bmatrix}
	\psi^{+ \dot 2} (z) \\ \psi^{- \dot 2} (z) 
	\end{bmatrix}
	=
 	\begin{bmatrix}
	e^{ i \phi_1(z)} \\  - e^{i \phi_2(z)}  
	\end{bmatrix} .
\label{Bosnpsi2real}
\ee
Every exponential should be understood to be normal-ordered (and we ignore cocycles). The stress-tensor (\ref{TJGasXpsiT}) can be written in the completely bosonic form,%
	\footnote{%
	Normal ordering of two operators $\scr A_1$ and $\scr A_2$ is defined by
	$$
	\lor \scr A_1(z) \scr A_2(z) \ror \equiv \lim_{\vare \to 0} \Big[ \scr A_1(z+ \vare) \scr A_2(z ) -  \langle  \scr A_1(z+ \vare) \scr A_2(z)  \rangle \Big] .	
	$$
	}
\be
T(z) = - \frac{1}{2} \lim_{z' \to z} 
		\Bigg[ \sum_{i = 1}^4 \frac{1}{2} \pa X_i(z) \pa X_i(z') 
			+ \sum_{r=1}^2 \pa \phi_r(z) \pa \phi_r(z') 
			+ \frac{c}{(z - z')^2} \Bigg] .
\label{StressTensoBosoz}
\ee

Bosons are assumed to be periodic, so e.g.~$X^{\dot A A}(e^{2\pi i}z)=X^{\dot A A}(z)$.
Fermions can have Neveu-Schwarz or Ramond boundary conditions on $\mathbb C$.
The Ramond sector has a collection of degenerate vacua with holomorphic dimension $h = \frac{c}{24} = \frac{1}{4}$, and different charges under the global and R-symmetry SU(2) groups. The set of Ramond vacua can be obtained from the NS vacuum by the action of spin fields, conveniently realized as exponentials, e.g.~for the SU(2) doublet $S^{\a}(z)$,
\begin{align}
S^\pm(z) &= e^{\pm \frac{i}{2} [ \phi_1(z) - \phi_2(z) ]} 	.\label{spinfildpm}
\end{align}

To construct the orbifold $(T^4)^N/S_N$, one makes $N$ copies of the free SCFT and identifies them under the action of $S_N$; more explicitly, we take the $N$-fold tensor product 
$(\otimes^N T^4)/S_N$, and label operators $\scr O_I$ in each copy by an index $I = 1, \cdots , N$.
The energy tensor becomes
\be
T(z) = - \frac{1}{2} \lim_{z' \to z} 
		\sum_{I = 1}^N
		\Bigg[ \sum_{i = 1}^4 \frac{1}{2} \pa X_{i I}(z) \pa X_{i I}(z') 
			+ \sum_{r=1}^2 \pa \phi_{r I}(z) \pa \phi_{r I}(z') 
			+ \frac{c}{(z - z')^2} \Bigg] .
\label{StressTensoBosozTotal}
\ee
and the total central charge is $c_{\rm{orb}} = N c = 6N$.

Permutations of the copies can be realized by the insertion of \emph{twist operators} $\s_g(z)$, $g \in S_N$, which give a representation of $S_N$, and act on  the other operators by  twisting their  boundary conditions \cite{Dixon:1986qv},
\be
\scr O_{I} (e^{2\pi i}z)\sigma_{g}(0) = \scr O_{ g(I)}(z)\sigma_{g}(0) .
\label{twist-bc}
\ee
We are going to consider only cyclic twists, which form the building blocks of the Hilbert space of the orbifold theory \cite{Dijkgraaf:1996xw}. So, denoting by $(n)$ a generic cycle of length $n$, we consider
$g = (1)^{N - n} (n) \cong (n)$,
leaving the trivial cycles implicit.
We denote by $\s_n$ the twist operator for a generic cycle of length $n$;
they cyclically permute the $n$ copies of the fields appearing in the cycle $(n)$, while leaving the remaining copies invariant.
The conjugacy class is represented by the orbit-invariant combination 
\be
\s_{[n]} \equiv \frac{1}{\scr S_n(N)}  \sum_{h \in S_N} \s_{ h^{-1}(n) h} 
\label{Normalsighat}
\ee
where the representing cycle can be taken to be $(n) = (1 \cdots n)$, and the combinatorial factor $\scr S_n(N)$ makes the two-point function normalized, i.e.
\be
\big\langle  \s_{[n]}(z_1, \bar z_1) \s_{[m]}(z_2, \bar z_2) \big\rangle = \frac{\delta_{mn}}{| z_1 - z_2|^{2 \Delta_n}} .
\ee
The well-known (total) conformal dimension of a twist $\s_n(z,\bar z)$ is \cite{Lunin:2001pw,Dixon:1986qv} 
\be\label{twistdim}
\Delta^\s_n = h^\s_n + \tilde h^\s_n, 
\qquad
h_n^\s = \frac{1}{4} \Big( n - \frac{1}{n} \Big) = \tilde h^\s_n .
\ee

The $n$-twisted Ramond sector is generated by twisted spin operators with the appropriate $\frac1n$ rescaling of their weights. For the representative permutation $(1\cdots n)$, the $n$-twisted R-charged fields $R^{\a}_n(z)$ are
\begin{align}
R_n^{\pm}(z) &\equiv \exp \left( \pm \frac{i}{2n} \sum_{I = 1}^n \big[ \phi_{1,I}(z) - \phi_{2,I}(z)  \big] \right) \s_{(1 \cdots n)}(z)	\label{Rampmnnoninv}
\end{align}
with a similar construction for the neutral $R^{\dot A}_n(z)$.
From these, we can compose $S_N$-invariant combinations $R^\zeta_{[n]}(z)$, $\zeta = \pm, \dot A$ by summing over orbits, as we did for the  normalized $S_N$-invariant twists $\s_{[n]}$.
For example, the R-charged fields, with which we will be primarily concerned, are written explicitly as
\be
R^\pm_{[n]}(z) \equiv 
	\frac{1}{\scr S_n(N)}
	\sum_{h \in S_N}
	\exp \left( \pm \frac{i}{2n} \textstyle\sum_{I=1}^n \big[ \phi_{1,h(I)}(z) - \phi_{2, h(I)}(z)  \big] \right) \s_{h^{-1} (1 \cdots n) h}(z)
	\label{Rampmninv}
\ee
where in the exponential we sum over $h(I) = \{ h(1) , \cdots, h(n)\}$, the image of the original copy set $I = \{1, \cdots ,n\}$ under the permutation $h$.
The $R^\pm_{[n]}$, like the spin fields $S^\pm$, form a doublet of  SU(2)$_L$ and a singlet of SU(2)$_1$, with charges 
$j^3=\pm \tfrac{1}{2}$ and $\frak j^3 = 0$.
On the other hand, the $R^{\dot A}_{[n]}(z)$ form a singlet of R-symmetry and a doublet of SU(2)$_1$, with charges $j^3 = 0$ and $\frak j^3 = \pm \frac{1}{2}$.
The conformal weight of the $R^{(\zeta)}_{n}$ is
\be
h^R_n = \frac{n}{4} ,	\label{hRn}
\ee
obtained from the combined weights of the exponential and the twist. Completely analogous fields $\tilde R^\zeta_{[n]}(\bar z)$, with dimension $\tilde h^R_n = h^R_n$, make the anti-holomorphic sector. 
The normalization factor $\scr S_n(N)$ ensures that the two-point functions are normalized, granted that the non-$S_N$-invariant functions are normalized:
\begin{align}
\big\langle R^\mp_{[n]}(\infty) \, R^\pm_{[n]}(0) \big\rangle
	=
	1
	=
\big\langle R^\mp_n(\infty) \, R^\pm_n(0) \big\rangle	.					\label{2ptRnonSn}
\end{align}


Let us examine these fields a little further.
The Hilbert space of the orbifold theory, $\cal H_{orb} = \oplus_{[g]} \cal H_g$, is a direct sum of sectors invariant under the conjugacy classes of $S_N$ \cite{Dijkgraaf:1996xw}. 
The latter are given by the irreducible decomposition of $g \in S_N$ into disjoint cycles, $[g] = \prod_{k \in \mathbb N} (k)^{q_k}$, with $\sum_k k q_k = N$, and we are interested in the simplest sector,
corresponding to $[g] = (1)^{N-n}(n)$. 
This Hilbert space, $\cal H_{(n)}$, is invariant under the centralizer subgroup 
$S_{N-n} \times \mathbb Z_n$, 
where $S_{N-n}$ permutes the $N-n$ trivial cycles $(1)$, and $\mathbb Z_n$ acts on the elements permuted by $(n)$.
It can be further decomposed as \cite{Dijkgraaf:1996xw}
$$
\cal H_{(n)} = \cal S^{(N - n)} \otimes \cal H^{\mathbb Z_n}_{(n)} ,
$$
where $\cal S^{(N - n)}$ is the symmetric tensor product of copies entering the trivial cycles, and where $\cal H^{\mathbb Z_n}_{(n)}$ corresponds to the copies permuted by $(n) \in \mathbb Z_n$.
States in $\cal H_{(n)}$ can be interpreted as a string with winding number $n$. 
We can think of the construction of the operators $R^{(\zeta)}_n$ as exciting the $n$-wound copies to the Ramond ground state, while leaving the unwound copies in the NS vacuum.  
Thus the operators $R^{(\zeta)}_n$ correspond to states
$
\ket{R^{(\zeta)}_n} = \ket{\varnothing}^\NS_{N-1} \otimes \ket{\varnothing , \zeta }^\R_{(n)}
$
where $\ket{\varnothing}^\NS_{N-1} \in \cal S^{N-1}$ is the NS vacuum of the non-twisted copies and 
$\ket{\varnothing , \zeta }^\R_{(n)}$ are Ramond ground states of a CFT defined on $\cal H^{\mathbb Z_n}_{(n)}$. Since the latter CFT involves $n$ copies of the $\cal N = 4$ SCFT,  it has central charge $c = 6n$; its Ramond ground states have the conformal weight $\frac{c}{24} = \frac{n}{4}$ in Eq.(\ref{hRn}).
We will sometimes refer to $R^\a_{[n]}$ as `Ramond ground states', but it should be kept in mind that this is an abuse of nomenclature, as $R^\a_n$ are ground states only of the $n$-wound string; the true Ramond ground states of the orbifold theory have conformal weight $h = \frac1{24} c_{orb} = \frac14 N$, and are made either by $R^\a_{[N]}$, i.e.~by single-cycle fields with maximal twist $n = N$, or, more generally, by ``composite fields'' $\prod_k(R^{\a_k}_{n_k})^{k}$ with $\sum_k k n_k = N$.

The main objective of this paper is to describe how the dimensions $h^R_n = \frac14 n$ are corrected when the free orbifolded SCFT is perturbed by a marginal operator.

\section{Away from the free orbifold}	\label{SectAwayOrbifold} \label{SectRenormConfData}

A marginal deformation of the free orbifold turns the theory into an interacting SCFT, with the action
\be
S_\inter = S_\free  + \la \int d^2z \, \Oint(z, \bar z)  \label{defpercft}
\ee
parameterized by a dimensionless deformation parameter $\la$. In the large-$N$ limit, in which we will be interested, the deformation parameter $\la$ should scale with $N$ in such a way that the 't Hooft coupling
$\la_* \equiv \la / \sqrt N $
is held fixed as $N \to \infty$; see \cite{Lunin:2000yv,Pakman:2009zz}.

The ``scalar modulus'' interaction operator $\Oint$ is marginal, with total conformal dimension $\Delta = h + \tilde h = 2$. This dimension should not change under renormalization. Also, $\Oint$ must be a singlet of R-symmetry, in order for $\cal N = (4,4)$ SUSY not to be broken. From the 20 deformation operators, which correspond to the 20 SUGRA moduli (see \cite{Avery:2010er}), we consider the $S_N$-invariant singlet
\be
O^{(\inter)}_{[2]}(z, \bar z) = \e_{A  B} G^{-  A}_{-\frac{1}{2}} \tilde G^{ \dot -  B}_{-\frac{1}{2}} O^{(0,0)}_{[2]}(z, \bar z) 	\label{DeformwithMo}
\ee
constructed as a descendent of the NS chiral field $O^{(0,0)}_{[2]}(z)$ with $h = \frac{1}{2} = j^3$.

\bigskip

Let us review a few key results in conformal perturbation theory used in the next sections; see for example \cite{Keller:2019yrr} for more detail. 
For a marginal perturbation, the two-point function
\be
\big\langle \scr O(z_1, \bar z_1)  \scr O (z_2, \bar z_2) \big\rangle_\la = |z_{12}|^{-2 \Delta_\la}	\label{twopinla}
\ee
of a neutral and hermitian (for simplicity) operator $\scr O$ is still fixed by conformal symmetry, hence the effect of the marginal perturbation has to be a change of its conformal dimension. The $\la$ expansion of the functional integral gives 
\be
\begin{split}
\big\langle \scr O(z_1, \bar z_1)  \scr O (z_2, \bar z_2) \big\rangle_\la = 
	\frac{1}{|z_{12}|^{2\Delta}} 
	\Bigg[ 1
		& + 2 \pi C \la  \log \frac{|z_{12}|}{\La}
\\	
		& + \pi \la^2 \log \frac{|z_{12}|}{\La} \int \! d^2 u \,  G(u, \bar u)  + \rm{O}(\la^3) \Bigg]  ,
\end{split}	\label{PErtuTwoPtOOsecla}
\ee
where 
absence of a $\la$-index (e.g.~in $\Delta$) always indicates evaluation in the free theory, and the objects in the r.h.s.~are defined as follows.
At first order, $C$ is the structure constant coming from the three-point function 
\be
C = \big\langle \scr O(\infty)  \Oint(1) \scr O (0) \big\rangle .	\label{CstrucCons}
\ee
At second order,
$G(u,\bar u)$ is the undetermined part of the four-point function in terms of the anharmonic ratio
$u \equiv (z_{12}z_{34})/( z_{13}z_{24})$,
\be
\big\langle \scr O(z_1, \bar z_1) \Oint(z_3, \bar z_3) \Oint (z_4 , \bar z_4) \scr O (z_2 , \bar z_2) \big\rangle %
	=  \frac{G(u, \bar u) }{ |z_{13}|^2 |z_{32}|^2|z_{12}|^{2\Delta-2}} ,
	\label{4ptgenG}
\ee
 and
$\La$ is a cutoff for the integral
\be
\int \! \frac{d^2 z_3 }{|z_{13}|^2 |z_{32}|^2 |z_{12}|^{2\Delta-2}} = \frac{2\pi}{|z_{12}|^{2\Delta}} \log \frac{|z_{12}|}{\La}  , \qquad\quad \La \ll 1 .
\ee
The $\log \La$ divergence in the two-point function requires the introduction of an appropriate regularization  and a corresponding renormalization of the field $\scr O$. The logarithmic form of the divergent terms indeed has the effect of changing the exponent of the renormalized two-point function, thus changing $\Delta$.
The operators we are interested in have a vanishing three-point function with $\Oint$, i.e.~$C = 0$. The  corrections in (\ref{PErtuTwoPtOOsecla}) therefore  start at second order in $\la$, and the renormalized field is
\begin{flalign}
&& &\scr O^{(ren)}(z,\bar z) = \La^{\frac{1}{2} \pi \la^2 J} \scr O(z,\bar z) , &&
\\
\text{where} && & J \equiv \int \! d^2u \, G(u,\bar u)	. &&	\label{JintDef}
\end{flalign}
We can see that
\begin{align*}
\big\langle \scr O^{(ren)}(z_1,\bar z_1)  \scr O^{(ren)} & (z_2,\bar z_2) \big\rangle_\la
	= 
\La^{\pi \la^2 J} \big\langle \scr O(z_1,\bar z_1) \scr O(z_2,\bar z_2) \big\rangle_\la
\\
	&= 
	\Big( 1 + \pi \la^2 J  \log \La + \cdots \Big)
	|z_{12}|^{-2\Delta}
	\Big(1 + \pi \la^2 J \log \frac{|z_{12}|}{\La} + \cdots \Big)
\\
	&= 
	\Big( 1 + \pi \la^2 J  \log |z_{12}| + \cdots \Big)
	|z_{12}|^{-2\Delta}
\\
	&= 
	|z_{12}|^{\pi \la^2 J}	|z_{12}|^{-2\Delta}
\end{align*}	
so the $\La$-divergence is canceled, and the free-theory dimension $\Delta$ has flowed to a $\la$-dependent value
\be
\Delta_\la = \Delta - \tfrac{\pi}{2} \la^2 J  + \rm{O}(\la^3) .	\label{DeltccorinG}
\ee

The integral $J$ also gives the \emph{first}-order  $\la$-correction to the particular structure constant in (\ref{CstrucCons}). This can be seen from the functional integral expansion of the corresponding three-point function.
For our case where the free-theory constant vanishes, we find
\be
C_\la =  \la J + \rm{O}(\la^2)  .	\label{deltaCstrc}
\ee

To compute the integral (\ref{JintDef}),  we need to be able to calculate the four-point function (\ref{4ptgenG}) in the free orbifold theory. 
In the next section, we show how to do this.

\section{Four-point functions}	\label{SectFourPointFuncts}

Our goal is to compute four-point functions%
	\footnote{%
	A note on convention: in this paper, fields inside correlation functions are to be understood in \emph{two}-dimensional theory, e.g.~$\s_n(z,\bar z)$, instead of $\s_n(z)$. However, when fixing a point in ${\mathbb C}^2$ we only write one argument for economy of notation. Thus, in (\ref{4pointR1Gu}), it should be understood that $\scr O_{[n]}(0) = \scr O_{[n]}(0,\bar 0)$, $\Oint(1) = \Oint(1,\bar 1)$, etc.
	}
\be
G(u, \bar u) = \big\langle \scr O_{[n]}^\dagger(\infty) \Oint(1) \Oint (u , \bar u) \scr O_{[n]}(0) \big\rangle 	\label{4pointR1Gu}
\ee
for primary operators $\scr O$ in the $n$-twisted sectors of the orbifold SCFT$_2$. 
Some aspects of the computation are universal, depending only on the nature of the twists:
we start by describing the covering surface appropriate to the twisted structure of (\ref{4pointR1Gu}); then we describe the stress-tensor method to compute the simplest four-point function with this structure, containing only bare twists. Finally, we turn to the cases containing interaction operators with $\scr O_{[n]}$ as the charged Ramond ground state, or as a bare twist field.

\subsection{The covering surface}	\label{SectCoveSurf}

Twisted correlators such as (\ref{4pointR1Gu}) are complicated functions, with specific monodromies of their arguments fixed by their (bare) twist fields constituents.
The standard way \cite{Lunin:2000yv} of implementing the boundary conditions (\ref{twist-bc}) for $G(u,\bar u)$ is to map the `base sphere' $S^2_{\rm{base}} = \mathbb C \cup \infty$ to a ramified `covering surface' $\Sigma_{\rm{cover}}$, whose ramification points correspond to the position and the order of twists operators.
At large $N$, the leading contribution comes from genus-zero covering surfaces. Denote coordinates on the base by $z \in S^2_{\rm{base}}$, and coordinates on the covering sphere by $t \in S^2_{\rm{cover}}$,
and fix the four punctures on each surface  to be
$$
\{z = 0 \} \mapsto \{t = 0\}, \{z = 1\} \mapsto \{t = t_1\}, \{z = u \} \mapsto \{t = x\}, \{z = \infty\} \mapsto \{t = \infty\}.
$$ 
The method for finding  $z(t)$ for generic monodromies was pioneered in \cite{Lunin:2000yv} and generalized in \cite{Pakman:2009zz}. 
For the specific monodromies (and topology) above,  
\be\label{cover}
z(t)=\left({t\over t_1}\right)^n \left( \frac{t-t_0}{t_1-t_0} \right) \left( \frac{t_1-t_\infty }{t-t_\infty} \right) .
\ee
The monodromies at $z = 0$ and $z= \infty$ are evident, but at $z = 1$ and $z = u$ they are implicit in the derivative $z'(t)$, which must vanish at every branching point.  Indeed,
\be
\frac{d z}{d t}  = \frac{t_1- t_\infty}{t_1(t_0 - t_1)} \frac{t^{n-1}}{(t - t_\infty)^2} 
				\Big[ ( t_\infty - t_0) t - n (t - t_0)(t - t_\infty) \Big]
	\label{derz}
\ee
vanishes at $t = 0$ with the correct monodromy, while $x$ and $t_1$ must be the roots of  the quadratic expression in brackets. 
This quadratic equation relates the parameters $t_1,t_0,t_\infty$ and $x$, 
\be
x + t_1 = \tfrac{n-1}{n} t_0 + \tfrac{n+1}{n} t_\infty ;
\qquad
x t_1 = t_0 t_\infty .	\label{sumprdquad}
\ee
We are free to choose one of the ratios $t_0/x$, $t_0/t_1$, $t_\infty / t_1$ and $t_\infty/x$ as long as they satisfy the two conditions (\ref{sumprdquad}), and we choose
\be
1 - \frac{t_0}{ x} = \frac{1}{x} , 
\quad
\text{hence}
\quad
1 - \frac{t_\infty }{ x} = \frac{1}{x+n},
\quad
1 - \frac{t_1}{x} = \frac{2x + n - 1}{x(n+x)} 
	\label{ArtFrolChoice}
\ee
which gives $u = z(x)$ as \cite{Arutyunov:1997gi,Arutyunov:1997gt}
\be
u(x)={x^{n-1}(x+n)^{n+1}\over (x-1)^{n+1}(x+n-1)^{n-1}} ,
	\label{ux}
\ee
a rational function.

\subsection{Four-point functions}	\label{SectStressTensrMth}

The covering map encodes the monodromies of functions like (\ref{4pointR1Gu}), with the twist structure
\be
g(u , \bar u) \equiv \big\langle \s_n(\infty) \s_2 (1) \s_2(u,\bar u) \s_n(0) \big\rangle ,	\label{ssss}
\ee
into the ramification points of the covering surface.
One way of computing $g(u,\bar u)$, formulated by Lunin and Mathur \cite{Lunin:2000yv}, is to cut circles around the ramification points, replace them with vacua and compute the functional integral directly. 
An alternative%
		\footnote{%
		Still other ways of computing general four-point functions $\langle \s_m \s_n \s_p \s_q \rangle$ have been recently given \cite{Roumpedakis:2018tdb,Dei:2019iym}.
		}
 \cite{Dixon:1986qv} is to use the 
conformal Ward identity: if one is able to find the residue $r(u)$ of the following function \emph{on the base},
\be\begin{split}
f(z) &= \frac{ \big\langle T(z) \s_n(\infty) \s_2(1) \s_2(u, \bar u) \s_n(0) \big\rangle
		}
		{ 
		\big\langle \s_n(\infty) \s_2(1) \s_2(u, \bar u) \s_n(0) \big\rangle
		}
	= \frac{h}{(z - u)^2} + \frac{r(u)}{z - u} + \text{non-sing.} ,
\end{split}\ee
the Ward identity gives a differential equation
\be\label{equS}
\pa_u \log g(u)= r(u) 
\ee
which can be solved for the holomorphic part of $g(u,\bar u) = g(u) \tilde g(\bar u)$.
The anti-holomorphic part $\tilde g(\bar u) = \bar g(\bar u)$ is obtained likewise, using $\tilde T(\bar z)$.

In simpler orbifold theories, it is possible to find  $r(u)$ by  engineering the function with the appropriate poles and monodromies \cite{Dixon:1986qv}. Here, we can follow Refs.\cite{Arutyunov:1997gi,Arutyunov:1997gt,Pakman:2009ab,Pakman:2009zz} and use the covering surface as an aid, by computing  the correlation functions on $S^2_{\rm{cover}}$, where the monodromies are trivial, and then mapping back: $f(z)$ is a function of the position of the stress tensor which, unlike the twists, is \emph{not} placed on a branching/ramification point --- hence mapping from covering to base is just a conformal transformation. On the covering, 
\be
f_{\rm{cover}}(t) = \frac{ \langle T(t) \mathds 1 \rangle}{ \langle \mathds 1 \rangle} = 0
\ee
because the twists disappear, and when mapping back to base only the anomalous transformation of $T$ does not cancel in the fraction, so
\be
f(z) = \sum_I \left[ \frac{c}{12} \{ t_I , z\} + \left( \frac{dt_I}{dz} \right)^2 f_{\rm{cover}}(t_I(z))  \right]
	= \sum_I \tfrac{1}{2} \{ t_I , z\} .
	\label{fzSc}
\ee
The position of the twists appear as parameters implicit in the inverse maps $z \mapsto t$, which encode the twist structure of (\ref{ssss}). 
There is a sum over $I$ in Eq.(\ref{fzSc}) because $T(z)$ is a sum over copies (\ref{StressTensoBosozTotal}). Around a branching point, there is one inverse map $t_I(z)$ for each copy entering the corresponding twist; at $z = u$, the insertion point of $\s_2$, there are two maps, which can be found locally \cite{Arutyunov:1997gt,Pakman:2009mi}, as follows.
Take the logarithm of the ratio $z(t) / z(x)$, i.e.~
$\log (z / u) = n \log \frac{t}{x} + \log \frac{t - t_0}{x - t_0} - \log \frac{t- t_\infty}{x - t_\infty}$,
and expand both sides, 
\be
\sum_{k = 1}^\infty b_k (z - u)^k = (t-x)^2 \sum_{k = 0}^\infty a_k (t -x)^k ,
\quad
\text{hence}
\quad
t - x = \sum_{k =1}^\infty c_k (z - u)^{k/2} .
\label{PoeSeritxzu}	
\ee
In the first equation, the coefficients are found from the Taylor expansions,
\begin{flalign}
&& 
&b_k = \frac{(-1)^{k+1}}{k u^k}  ,
\quad
a_k	= \frac{(-1)^{k+1}}{k+2} \left[ \frac{1}{(x - t_0 )^{k+2}} - \frac{1}{(x - t_\infty )^{k+2}} + \frac{n}{x^{k+2}} \right] . &&
	\label{PoeSeritx}
\end{flalign}
The coefficients $c_k$ are solved in terms of $a_k$ and $b_k$ order by order, by inserting the $c_k$ power series into the first equation in (\ref{PoeSeritxzu}).
The multiple inverses $z \mapsto t$ appear as multiple solutions for the $c_k$.
After solving for the $c_k$, we can put the powers series into the r.h.s.~of Eq.(\ref{fzSc}), expand to order $(z - u)^{-1}$ and extract the desired residue. The coefficients $c_1$, $c_2$ and $c_3$ completely determine the result up to this order, 
\begin{align}
c_1 = \pm \sqrt{ \frac{b_1}{a_0}} ,
\quad
c_2 = -\frac{a_1 b_1}{2 a_0^2},
\quad
c_3 = \pm \frac{4 a_0 a_2 b_1^2 - 4 a_0^3 b_2  - 5 a_1^2 b_1^2 }{ 8 a_0^{7/2} \sqrt{b_1}} .
	\label{txSerzucof}
\end{align}
As expected, there are two solutions.
When the parameters  $t_0$ and $t_\infty$ in $a_k,b_k$ are written explicitly in terms of $x$,  these coefficients are functions of $x$ alone, thus we find the residue $r$ as a function of $x$. One can check that $r(x)$ is the same for both choices of the $c_k$.

Solving Eq.(\ref{equS}) requires expressing $r(x)$ as an explicit function of $u$, but there are multiple inverses of $u(x)$. It is easier to make a change of variables, and solve instead the differential equation
\be
\pa_x \log g(x) = u'(x) r(x)  ,	\label{diffeqGx}
\ee
whose solution is
\be
g(x)  = c_\s
	 \frac{
		x^{ - \frac{2+5n(n-1)}{8n} }
		(x-1)^{ \frac{ 2 + 5n(n+1) }{ 8n } }
		(x+n)^{ \frac{ 2 - n(n+1) }{8n}}
		(x+n-1)^{- \frac{ 2 - n(n-1) }{8n} }
		}{
		(x+\frac{n-1}{2})^{1/4}
		}	.
\label{functwists}\ee
The integration constant $c_\s$ has to be determined by looking at OPE limits (see App.\ref{SectStrucConst}).%
		\footnote{%
		We emphasize that the function $\langle \s_n(\infty) \s_2(1) \s_2(u,\bar u) \s_n(0)\rangle$ is known in the literature, calculated by other methods. Our point is to take it as an instructive example of the specific method we use. 	}

Now, we have found a function parameterized by the pre-image of $u$ under the covering map $z(t)$. For fixed $u = u_*$, there are $\bf H$ different pre-images $x_{\frak a}$, $\frak a = \frak1, \cdots, \bf H$, solutions of the equation
\be
 x^{n-1}(x+n)^{n+1} - u_* (x-1)^{n+1}(x+n-1)^{n-1} = 0 .	\label{RamifEqHu}
\ee
The degree of the polynomial shows that ${\bf H}  = 2n$.
Note that this is not the number of sheets of the ramified covering ($u$ is the position of a branching point), it is the number of \emph{different covering maps} with the assumed monodromy conditions; $\bf H$ is a Hurwitz number \cite{Pakman:2009zz,Pakman:2009ab,Pakman:2009mi}.

The method has thus yielded $\bf H$ functions $g(x_{\frak a}(u))$.
This was expected, because the $S_N$ structure of the composition of cycles in Eq.(\ref{ssss}) is not completely fixed. 
Labeling cycles by the position of their twists operators, those entering $g(u,\bar u)$ must compose to the identity,
\be
(n)_\infty (2)_1 (2)_u (n)_0 = 1	,	\label{compto1}
\ee
otherwise the correlator vanishes.
There are several collections $\{(n)_\infty, (2)_1 , (2)_u , (n)_0 \}$ of cycles which solve Eq.(\ref{compto1}),%
		\footnote{%
		 The total number of such solutions can be found with Frobenius' formula 
		 \cite{lando2013graphs}.
		 }
and these collections can be arranged into equivalence classes defined by
\be
(n)_\infty (2)_1 (2)_u (n)_0 
\sim
h (n)_\infty h^{-1} \, h (2)_1 h^{-1} \  h (2)_u h^{-1} \, h (n)_0 h^{-1}
\quad
\forall
\quad h \in S_N .
\ee
The existence of different such equivalence classes is the reason for the existence of different functions $g(x_{\frak a}(u))$;  there are precisely ${\bf H} = 2n$ equivalence classes \cite{Pakman:2009zz}.
Inside each of these classes, let $C_{\bf s}(N)$ be the number of collections $\{(n)_\infty, (2)_1 , (2)_u , (n)_0 \}$ for which the cycles involve a fixed number ${\bf s}$ of distinct elements of $\{1,2,\cdots, N\}$. Then it can be shown \cite{Pakman:2009zz} that $C_{\bf s}$ is the same for all classes, and that, for large $N$, it scales as 
\be
C_{\bf s} = N^{{\bf s} - \frac{1}{2} \sum_{r=1}^4 n_r } \left[ \varpi(n_r)  +  \rm{O}(1/N) \right] ,	\label{CsNlim}
\ee
where $n_1 = n = n_4$ and $n_2 = 2 = n_3$ are the order of the $q = 4$ twists involved in (\ref{ssss}).
But $n_r - 1$ is also the order of the ramification points of the covering surface,  $\bf s$ is the number of its sheets, hence its genus is fixed by the Riemann-Hurwitz formula
\be
 {\bf g} = 1 - {\bf s} + \frac{1}{2} \sum_{r=1}^q (n_r - 1).
  \label{RiemHurwForm}
 \ee
We thus see that 
$
C_{{\bf s}} (N) \equiv  C_{{\bf g}} (N)	\sim N^{- {\bf g} -  1} ,
$
therefore the covering surface with ${\bf g} = 0$ constructed in \S\ref{SectCoveSurf} gives the leading contribution at large $N$ \cite{Lunin:2000yv}.
For our four-point functions, the Riemann-Hurwitz formula gives ${\bf s} = - {\bf g} + n+1$, hence we see that, for the covering surface to have genus zero, we must have  
\be
1 \leq n < N .	\label{2nN}
\ee
When we sum over the orbits of individual cycles to make an $S_N$-invariant correlation function, we get all terms in each of the equivalence classes above, 
\be
\big\langle \s_{[n]}(\infty) \s_{[2]} (1) \s_{[2]}(u,\bar u) \s_{[n]}(0) \big\rangle
	= \frac{\varpi(n)}{N} \sum_{\frak a = \frak1}^{\bf H} g( x_{\frak a}(u)) \bar g(\bar x_{\frak a}(\bar u)) . 
	\label{foupoinGuuxj}
\ee
This sum corresponds to different OPE channels resulting from composing the twist permutations, not only for $g(u,\bar u)$ but for the other functions $G(u,\bar u)$ which share the same twist structure.

\subsubsection{Charged Ramond fields}

Let us now turn to the function 
\be
G_R(u, \bar u) = \big\langle  R^-_{[n]}(\infty) \Oint(1) \Oint (u , \bar u) R^+_{[n]}(0) \big\rangle .	\label{4pointR1GuRamon}
\ee
The Ramond fields $R^\pm_{[n]}(z,\bar z)$ are lifted to the corresponding spin field $S^\pm(t,\bar t)$, so we compute
\be\label{methodcov}
F_{\rm{cover}} (t) =
	 \frac{\big\langle T(t) S^-(\infty) \Oincov(t_1,\bar t_1) \Oincov (x , \bar x) S^+(0) \big\rangle}{\big\langle S^-(\infty) \Oincov(t_1, \bar t_1) \Oincov(x , \bar x) S^+(0)  \big\rangle} 
\ee
and then find the residue $H$ of the function
\be
\begin{split}
F (z) &= 	 \frac{\big\langle T(z) R^-_{[n]} (\infty) \Oint(1) \Oint (u , \bar u) R^+_{[n]}(0) \big\rangle}{\big\langle R^-_{[n]}(\infty) \Oint(1) \Oint (u , \bar u) R^+_{[n]}(0) \big\rangle} 
\\
	&= 2 \left[ \frac{1}{2} \big\{ t , z \big \} + \left(\frac{dt}{dz}\right)^2  F_{\rm{cover}}(t(z), x) \right] 
	=  \frac{H(x)}{z - u} + \cdots
\end{split}		\label{FzusumovaT}
\ee
with $t(z)$ one of the maps obtained from Eqs.(\ref{PoeSeritxzu}) and (\ref{txSerzucof}).

The deformation operator, denoted by $\Oincov(t,\bar t)$ --- without a twist index since there are no twists on the covering surface --- can be expressed on $S^2_{\rm{cover}}$  in terms of the basic fields only, because the contour integrals in the super-current modes $G^{\a A}_{- \frac{1}{2}} = \frac{1}{2\pi i} \oint dz   G^{\a A}(z)$ just pick up a residue  (see e.g.~\cite{Burrington:2012yq}). The result is a sum of products of bosonic currents, free fermions and spin fields coming from the lifting of the NS chiral field
$O^{(0,0)}_{[2]} (z,\bar z) \mapsto S^+ (t)\tilde S^{\dot +}(\bar t)$. 
Writing spin fields as exponentials,
\be
\begin{split}
 \Oincov  =
		a_{\rm{int}} \Big[
		& \lor \pa X^{\dot 1 1} \, e^{+\frac{i}{2} (\phi_1 + \phi_2)} 
			\left( \bar \pa X^{\dot 1 2} e^{+\frac{i}{2} (\tilde \phi_1 + \tilde \phi_2)} 
				- (\bar \pa X^{\dot 1 1})^\dagger e^{- \frac{i}{2} (\tilde \phi_1 + \tilde \phi_2)} 
			\right) \ror	
\\
		- &	\lor \pa X^{\dot 1 2} \, e^{+\frac{i}{2} (\phi_1 + \phi_2)} 
			\left( (\bar \pa X^{\dot 1 2})^\dagger e^{- \frac{i}{2} (\tilde \phi_1 + \tilde \phi_2)} 
				+ \bar \pa X^{\dot 1 1} e^{+ \frac{i}{2} (\tilde \phi_1 + \tilde \phi_2)} 
			\right) \ror	
\\
		+ &	\lor (\pa X^{\dot 1 1})^\dagger e^{- \frac{i}{2} (\phi_1 + \phi_2)} 
			\left( (\bar \pa X^{\dot 1 2})^\dagger e^{-\frac{i}{2} (\tilde \phi_1 + \tilde \phi_2)} 
				+ \bar \pa X^{\dot 1 1} e^{+\frac{i}{2} (\tilde \phi_1 + \tilde \phi_2)} 
			\right) \ror	
\\
		+ &	\lor (\pa X^{\dot 1 2})^\dagger \, e^{-\frac{i}{2} (\phi_1 + \phi_2)} 
			\left( \bar \pa X^{\dot 1 2} e^{+ \frac{i}{2} (\tilde \phi_1 + \tilde \phi_2)} 
				- (\bar \pa X^{\dot 1 1})^\dagger e^{- \frac{i}{2} (\tilde \phi_1 + \tilde \phi_2)} 
			\right) \ror	
			\Big] .
\end{split}	\label{InteraOpera}
\ee  
The constant $a_{\rm{int}}$ can be conveniently chosen by a redefinition of the deformation parameter $\la$. For now, we leave it unspecified.
To compute the correlators, the strategy is to show that contractions of $T(t)$ with the fields in the numerator of (\ref{methodcov}) are always proportional to 
$
\cal G = \langle S^{-} (\infty) \Oincov(t_1, \bar t_1) \Oincov(x , \bar x) S^+(0)  \rangle
$,
appearing in the denominator of Eq.(\ref{methodcov}).
We can decompose $T(t) = T_B(t) + T_F(t)$ into bosonic and fermionic parts, respectively
\begin{align*}
T_B(t) &= - \tfrac{1}{4} \e_{\dot A \dot B} \e_{AB} \colon\! \pa X^{\dot A A}(t) \pa X^{\dot B B}(t) \colon
\\
T_F(t) &= - \tfrac{1}{2} \colon \!\! \big[ \pa \phi_1(t) \pa \phi_1(t) + \pa \phi_2(t) \pa \phi_2(t) \big] \! \colon 
\end{align*}
As far as bosons are concerned, each term of the product $\Oincov(t_1) \Oincov(x)$ has the structure
$\pa X^{\dot C C}(t_1) \pa X^{\dot E E}(x)$ multiplied by ``transparent'' fermionic or anti-holomorphic factors. Using the conformal Ward identity and the two-point functions (\ref{2ptFuncpsipsi}),
\begin{align*}
\big\langle T_B(t) \pa X^{\dot C C}(t_1) \pa X^{\dot E E}(x) \big\rangle
	&= \left[ \frac{\pa_{t_1}}{t - t_1}  + \frac{\pa_x}{t - x}  + \frac{1}{(t - t_1)^2} + \frac{1}{(t - x)^2} \right] \frac{2 \e^{\dot C \dot E} \e^{CE}}{(t_1 - x)^2}
\\
	&=  \frac{(t_1 - x)^2}{(t-t_1)^2 (t - x)^2}  \ \big\langle \pa X^{\dot C C}(t_1) \pa X^{\dot E E}(x)  \big\rangle .
\end{align*}
Hence we can recompose $\cal G$, and obtain
\be
\frac{\big\langle T_B(t) S^-(\infty) \Oincov(t_1, \bar t_1) \Oincov (x , \bar x) S^+(0) \big\rangle
	}{
\big\langle S^{-} (\infty) \Oincov(t_1, \bar t_1) \Oincov(x , \bar x) S^+(0)  \big\rangle	} 
	=
	\frac{(t_1 - x)^2}{(t-t_1)^2 (t - x)^2} .
	\label{TbROORcov}
\ee

For the fermionic part of the calculation, it is very helpful to organize $\Oincov$ as
\bsub
\begin{flalign}
&& \Oincov (t , \bar t) &\equiv  V_- (t , \bar t) + V_+ (t , \bar t) , &&
\\
\text{where} &&
V_+(t,\bar t) &= \Big[ 
		\big( a.h.)_{\dot 1 1} \pa X^{\dot 11} 
		-  
		\big( a.h.)_{\dot 1 2} \pa X^{\dot 12} 
		\Big] \lor e^{+ \frac{i}{2} ( \phi_1 +  \phi_2 )} \ror
\\
&& V_- (t,\bar t) &= \Big[ 
		\big( {a.h.})_{\dot 1 1\dagger} (\pa X^{\dot 11})^\dagger 
		+  
		\big( {a.h.})_{\dot 1 2\dagger} (\pa X^{\dot 12})^\dagger 
		\Big] \lor e^{- \frac{i}{2} ( \phi_1 +  \phi_2 )} \ror
\end{flalign}\label{DefofVpm}\esub
the $(a.h)$s being combinations of anti-holomorphic fields which can be read from (\ref{InteraOpera}). 
This makes it is clear that contractions with $\Oint$ are very simple, and
\begin{align}
&\textstyle\sum_r \lim_{v \to t} \big\langle \pa\phi_r(v) \pa \phi_r(t) S^-(\infty) \Oincov(t_1) \Oincov (x , \bar x) S^+(0) \big\rangle - \textstyle\sum_r \lim \contraction{}{\phi}{_r(t)}{\phi} \phi_r(t) \phi_r(v)
\nonumber \\
	 & = \Bigg[ \frac{(i/2)^2}{t^2} + \frac{(i/2)^2}{(t-t_1)^2} + \frac{(i/2)^2}{(t-x)^2} \Bigg]  \cal G
\nonumber \\
	&\quad +  \frac{2(i/2)^2}{(t - t_1)(t-x)} \big\langle S^-(\infty) \big[ V_-(t_1) - V_+(t_1) \big] [ V_-(x) - V_+(x) \big] S^+(0) \big\rangle 
\label{limsumOOVWDas}
\end{align}
The second line in the r.h.s.~can be further simplified because, since the only non-vanishing two-point functions (\ref{twopntboconj}) are between a field and its conjugate, it follows that
$\langle V_\pm (t,\bar t) V_\pm(v,\bar v) \rangle = 0$,
hence
\be
\big\langle S^-(\infty) \big[ V_-(t_1) - V_+(t_1) \big] [ V_-(x) - V_+(x) \big] S^+(0) \big\rangle 
= -  \cal G.
	\label{VpVmsimpl}
\ee
Putting this back in (\ref{limsumOOVWDas}),  $\cal G$ appears as a common factor canceled in (\ref{methodcov}), 
\be
\begin{split}
&\frac{
	\big\langle T_F(t) S^-(\infty) \Oincov(t_1,\bar t_1) \Oincov (x , \bar x) S^+(0) \big\rangle
	}{
\big\langle S^{-} (\infty) \Oincov(t_1, \bar t_1) \Oincov(x , \bar x) S^+(0)  \big\rangle	 
	}
	=
	 \frac{1}{4} \left[ \frac{1}{t^2} + \left( \frac{1}{ t  - t_1 } -   \frac{1}{ t - x } \right)^2 \right]  .
\end{split}		\label{TfROORcov}
\ee
Combining (\ref{TbROORcov}) and (\ref{TfROORcov}), we get
\be
\begin{split}
F_{\rm{cover}} (t) &= \frac{(t_1 - x)^2}{( t - t_1 )^2 (t - x)^2} 
		 + \frac{1}{4} \left[ \frac{1}{t^2} + \left( \frac{1}{t-t_1} - \frac{1}{t-x} \right)^2 \right] .
\end{split}		\label{FcoverRamond}
\ee
Inverting the maps, we find the residue $H(x)$ of $F(z)$ to be
\be
\begin{split}
H(x) = - \Big[ 
		&16 x^4 + 32 (2 n-1) x^3 
\\		
&		+ 4 (2 n-1) (10 n-7) x^2  
\\		
&		+ 4 (n-1) \left[10 (n-1) n+3\right] x
\\
&
		 +5 (n-2) (n-1)^2 n \Big] \big[ 4  n (n+2 x-1)^3 \big]^{-1}  .
\end{split}		\label{Hofx}
\ee
The solution of the differential equation 
$\pa_x \log G_R(x) = u'(x) H(x)$ 
is now easily found, 
\be
G_R(x) = C_R \; \frac {x^{\frac{5(2-n)}{4}}(x-1)^{\frac{5(2+n)}{4}}(x+n)^{\frac{2-3n}{4}}(x+n-1)^{\frac{2+3n}{4}}}{(x+\frac{n-1}{2})^4} .\label{func}
\ee
where $C_R$ is an integration constant.

\subsubsection{Bare twists}

Let us also consider 
\be
G_\s(u,\bar u) = 
\big\langle \s_{[n]}(\infty) \Oint(1) \Oint (u , \bar u) \s_{[n]}(0) \big\rangle  	\label{4pointSigGu}
\ee
appearing in the second-order correction of the two-point function of bare twist fields.
The computation of 
\be
\begin{split}
F_{\rm{cover}} (t) 
	&= 
	\frac{\big\langle T(t) \mathds 1(\infty) \Oincov(t_1,\bar t_1) \Oincov (x , \bar x) \mathds 1(0) \big\rangle}{\big\langle \mathds 1(\infty) \Oincov(t_1,\bar t_1) \Oincov (x , \bar x) \mathds 1(0)  \big\rangle}
\\	
	&= \frac{(t_1 - x)^2}{( t - t_1 )^2 (t - x)^2} 
		 + \frac{1}{4} \left( \frac{1}{t-t_1} - \frac{1}{t-x} \right)^2  ,
\end{split}		\label{Fcoversigma}
\ee
goes as before (but is simpler), and  we find
\be
G_\sigma(x) = C_\sigma \; 
		\frac{
		x^{ - \frac{1 -10n + 5n^2 }{ 4n } }
		(x-1)^{ \frac{1+ 10 n + 5n^2 }{ 4n } }
		(x+n)^{ \frac{1 + 2n - 3n^2 }{ 4n } }
		(x+n-1)^{ - \frac{1 - 2n  - 3n^2 }{ 4n } } 
		}{
		(x+\frac{n-1}{2})^4
		} .\label{funcsig}
\ee
where $C_\sigma$ is an integration constant.
The same function has been computed in App.E of Ref.\cite{Pakman:2009mi}, but using a different parameterization map $u(x)$, in place of (\ref{ux}) (hence their function $G(x)$ is different from ours).

\section{OPE limits, fusion rules and structure constants}		\label{SectOPElimits}

The short-distance behavior of $G(u, \bar u)$ in the limits  $u \to 1,0,\infty$ contains the complete conformal data of the operator product expansions of the fields involved  --- i.e.~the OPE fusion rules.
Recall that super-conformal invariance fixes the form of the OPE algebra of generic primary holomorphic fields $\scr O^{(j^3_k)}_k(u)$ with dimensions $\Delta_k$ and  R-charges $j^3_k$ to be
\be
 \scr O_1^{(j^3_1)} (u,\bar u) \scr O_2^{(j^3_2)}(0) = \sum_k  C_{12k} \, |u|^{\Delta_k - \Delta_1 - \Delta_2} \scr O_k^{(j^3_k)}(0) + \text{descendants},
 \label{OPE}
 \ee
with structure constants $C_{12k}$, and $j^3_k = j^3_1+ j^3_2$.

\subsection{The OPE of two interaction operators}

The OPE of two interaction operators appears in the limit $u \to 1$ of $G(u,\bar u)$.
To extract  this limit from  $G(x)$, we have to find the inverse maps $x_{\frak a}(u)$ which contribute to the singularities near $u = 1$. 
For both $G_R(x)$ and $G_\sigma(x)$,
there are clearly only two contributions, i.e.~limits where $G(x)$ becomes singular, namely:%
		\footnote{%
		These correspond to the solutions of $t_1(x) = x$. Fortunately, we do not need to find the other solutions of the $2n$th-order polynomial equation (\ref{RamifEqHu}) for $u_* =1$.}
 $x = \infty$ and $x = \frac{1-n}{2}$, the former with multiplicity one, and the latter with multiplicity three. We label the two corresponding functions, given in (\ref{xaxp4ns}), as $x^1_{\frak a}(u)$, with a (gothic) index $\frak a = \frak{1,2}$, and the superscript indicating that $u \to 1$.
Each function gives a channel of the fusion rule, according to Eq.(\ref{foupoinGuuxj}). 
Both functions $G_R(x)$ and $G_\sigma(x)$ have the same behavior in these limits, as it was necessary for consistency, since both functions should give the same OPE
$
[ \Oint ] \times [ \Oint ] = [ \mathds 1] + [  \s_{[3]} ] ,	
$
where the r.h.s.~is based on the composition of permutations. 
We mostly focus on $G_R(x)$ in what follows, similar calculations for $G_\s(x)$ are listed in Appendix \ref{OPEsss}.

\vspace{3mm}

\noindent
\textbf{Determining the constants of integration}

\vspace{1mm}

\noindent
For $x \to \infty$,   $G_R(x) \approx C_R x^2$. 
Inserting $x = x_{\frak1}^1(u)$ given by Eq.(\ref{xaxp4ns}), we obtain
\be
G_R (x_{\frak1}^1(u)) = C_R \, {16n^2\over (1-u)^2} + 0 \times {1\over 1-u} + \text{non-singular}	\label{Gxinfu}
\ee
By formula (\ref{OPE}), since $\Oint$ has weight $h_{int} =1$, the leading singular term shows an operator of dimension $h = 2 - 2 = 0$ --- the identity operator. 
Also, the coefficient  of the term $\sim (1-u)^{-1}$ is zero, hence there is no contribution from a field of  dimension  $h = 1$, as it was to be expected for a truly marginal deformation.

\bigskip

The function in Eq.(\ref{Gxinfu}) corresponds to a correlator where the permutations in the twists form one representative element of the equivalence class where the 2-cycles of the interaction operators cancel. This happens when they share both elements. At order $N^{-1}$, there must be ${\bf s} = n+1$ elements entering the permutation, c.f.~Eq.(\ref{CsNlim}), so we can take this representative function to be
\be
\big\langle R^-_{(1,\cdots,n)} (\infty) \Oincov_{(1,n+1)}(1) \Oincov_{(1,n+1)}(u,\bar u) R^+_{(n, \cdots, 1)} (0) \big\rangle = G_R( x^1_{\frak1}(u)) G_R(x^1_{\frak1}(\bar u))	\label{fornonSGu1}
\ee
or any other with a global relabeling of elements in the cycles.

We now fix the constant $a_{\rm{int}}$ in (\ref{InteraOpera}) so that the non-$S_N$-invariant two-point functions are normalized,
\be
\big\langle \Oincov_2(\infty) \Oincov_2(1) \big\rangle = 1 .	\label{normOint}
\ee
Note that in these functions the two-cycles must share both of their elements, since, as in Eq.(\ref{compto1}), we must have $(2)_\infty (2)_1 = 1$. 
With this definition, the normalized $S_N$-invariant operator is
\be
\Oint (z,\bar z) = \frac{1}{\scr S_2(N)} \sum_{h \in S_N} \Oincov_{h^{-1} (12) h}(z,\bar z) .
\ee
Together with the normalization (\ref{2ptRnonSn}), inserting the limit (\ref{Gxinfu}), back into the four-point function (\ref{fornonSGu1}) we find $16n^2 C_R = 1$.
The same reasoning can be applied to the function $G_\s(x)$, which has the exact same limit as (\ref{Gxinfu}) in this channel.
Therefore
\be
C_R = \frac{1}{16n^2} = C_\sigma .		\label{CRCs}
\ee
With the functions $G(x)$ completely fixed, we can now look at other OPEs and derive structure constants.

\vspace{3mm}

\noindent
\textbf{The $\s_{3}$ channel}

\vspace{1mm}

\noindent
In the other channel corresponding to $u \to 1$, we must expand $G_R (x)$ around $x = \frac{1-n}{2}$, and insert $x^1_{\frak2}(u)$ given by Eq.(\ref{xaxp4ns}),
\be
\begin{split}
G_R(x^1_{\frak2}(u)) = & - \frac{4  (n^2 -1)^{\frac{1}{3}} \left( \frac{n+1}{n-1} \right)^{\frac{n}{2}} \left( \tfrac{1}{3} n \right)^{\frac{4}{3}} C_R}{(1-u)^{4/3}} 
\\	
	&
		 + \frac{\tfrac{1}{5} (7 + 2n^2) \left(\frac{n^2}{9(n^2 -1)}\right)^{\frac{1}{3} } \left( \frac{n+1}{n-1} \right)^{\frac{n}{2}}  C_R}{(1-u)^{2/3}}
\\	
	&
		- \frac{4 \cdot 3^{-\frac{7}{3}} \left( \frac{n+1}{n-1} \right)^{\frac{n}{2}} (n^2 -1)^{\frac{1}{3}} n^4 C_R}{(1-u)^{1 / 3}} + \text{non-singular}
\end{split}	\label{Channu1x1n2}
\ee
Once again, the coefficient of next-to-leading divergence, $\sim(1-u)^{3/3}$, vanishes, showing that there is no dimension-one operator in this conformal family either.
The leading singularity shows the presence of an operator of dimension 
$\tfrac{2}{3} = h^\s_3 $,
so we have found $\s_3$ itself, and the OPE
\be
\Oincov_2 (u,\bar u) \Oincov_2(1) = \frac{\big\langle \Oincov_2(\infty) \s_3(1) \Oincov_2(0) \big\rangle}{|1-u|^{8/3}} \, \s_3(1) + \cdots	\label{OPEOinOins3}
\ee	
whose structure constant is given in Eq.(\ref{Csint232}), and found independently from $G_\s(x)$.
Inserting the OPE into the correlation function we find the structure constant
\be
C^{R^- \s R^+} _{n3n} \equiv \big\langle R^-_n(\infty) \s_3(1) R^+_n(0) \big\rangle 
\ee
involving non-$S_N$-invariant Ramond fields and one three-twist.
The leading term in Eq.(\ref{Channu1x1n2}) gives us
\be
\log C^{R^- \s R^+} _{n3n} 
	=
	\left( n + \tfrac{2}{3} \right) \log (n+1) - \left(n - \tfrac{2}{3} \right) \log (n-1) - \tfrac{4}{3} \log n + \tfrac{4}{3} \log3 + \tfrac{1}{3}\log2 
\ee
after taking Eq.(\ref{Csint232}) into account.

\bigskip

The correlation function that gives Eq.(\ref{Channu1x1n2}) lies in an equivalence class where the 2-cycles of the interaction operator share only one element, thus forming $\s_3$. 
The multiplicity 3 of the solution $x = \frac{1-n}{2}$ for Eq.(\ref{RamifEqHu}) implies there are three different equivalence classes with this property. Representative functions for each of those classes are%
	\footnote{%
	An elegant and useful way of describing the different classes of permutations with the correct cycle structure and which satisfy Eq.(\ref{compto1}) is given in Refs.\cite{Pakman:2009zz,Pakman:2009mi} in terms of inequivalent diagrams.
	The permutations in Eqs.(\ref{ClassDiaga})-(\ref{ClassDiagc}) correspond, respectively, to the following diagrams in Ref.\cite{Pakman:2009mi}:
	a) the top diagram of Fig.4;
	b) the second diagram in Fig.4;
	c) the top diagram of Fig.5.
	}
\bsub\begin{align}
&G_R( x^1_{\frak2}(u)) G_R(x^1_{\frak2}(\bar u)) \nonumber
\\
&\qquad 
= \big\langle R^-_{(1,2,\cdots,n-1,n)} (\infty) \Oincov_{(1,n+1)}(u,\bar u) \Oincov_{(1,2)}(1) R^+_{(n+1, n, n-1, \cdots, 2)} (0) \big\rangle	\label{ClassDiaga}
\\
&\qquad 
= \big\langle R^-_{(1,2,\cdots,n-1,n)} (\infty) \Oincov_{(2,n+1)}(u,\bar u) \Oincov_{(1,n+1)}(1) R^+_{(n+1, n, n-1, \cdots, 2)} (0) \big\rangle
\\
&\qquad 
= \big\langle R^-_{(1,2,\cdots,n-1,n)} (\infty) \Oincov_{(1,2)}(u,\bar u) \Oincov_{(2,n+1)}(1) R^+_{(n+1, n, n-1, \cdots, 2)} (0) \big\rangle \label{ClassDiagc}
\end{align}\label{fornonSGu2}\esub
One can check that the permutations do satisfy Eq.(\ref{compto1}). 
Note that, by necessity, the twists in the Ramond fields are not the inverse of one another, so  the two-point function
\be
\big\langle R^-_{(1,2,\cdots,n-1,n)} (\infty) R^+_{(n+1, n, n-1, \cdots, 2)} (0) \big\rangle = 0 .
	\label{corRRs30}
\ee
Thus we see that the $\s_3$ channel of the fusion $[\Oint] \times [\Oint]$ is always present, because the interaction operator is necessarily an $S_N$-invariant object, but Eqs.(\ref{fornonSGu2}) and (\ref{corRRs30}) mean that $\s_3$ does not contribute to the correction of the two-point functions of \emph{individual}, non-$S_N$-invariant Ramond fields $R^\pm_n$.
It only contributes to the $S_N$-invariant combination $R^\pm_{[n]}$, by weaving together different individual terms.

\vspace{3mm}

\noindent
\textbf{The OPE $R^-_{n} R^+_{n}$}

\vspace{1mm}

\noindent
Although the positions of the Ramond fields are fixed in (\ref{foupoinGuuxj}), we can loosen the punctures back to Eq.(\ref{4ptgenG}), fix them differently as
$z_2 = \infty$, $z_3 = 0$, $z_4 = 1$, in which case $z_1 = u$, to find 
\be
\big\langle  \Oint(\infty) R^-_{[n]} (u,\bar u ) \Oint (0) R^+_{[n]}(1) \big\rangle 
			=  |1 -u|^{4 - n} G_R ( u, \bar u) .	
			\label{mastROORuinf2}
\ee
Now the limit $u \to 1$ corresponds to the OPE $R^-_{[n]} (u,\bar u) R^+_{[n]}(1)$.
The expansion near $u = 1$ for channel (\ref{Gxinfu}) is 
\be
\big\langle  \Oincov_2(\infty) R^-_{n} (u,\bar u ) \Oincov_2 (0) R^+_{n}(1) \big\rangle 
			=  \frac{1}{|1 -u|^n} + \cdots  	\label{mastROORuin3}
\ee
This corresponds to an operator of dimension zero, and is in fact the correct expression for the two-point function of Ramond fields, Eq.(\ref{2ptRnonSn}). 
In the channel (\ref{Channu1x1n2}) we now find the behavior $\sim (1-u)^{-n + \frac{8}{3} }$, indicating a twist-three operator of holomorphic weight 
\be
h = \frac{n + 4}{2} + h^\s_3.
\ee 

To understand the appearance of $\s_3$ in a channel of the OPE $R^-_n R^+_n$, let us consider the simpler case of the correlator with bare twists only.
Changing the points of Eq.(\ref{ssss}), we can find the OPE $\s_n \s_n$ from the limit $u \to 1$ of the function
\be
\big\langle \s_2(\infty) \s_n (1) \s_n(u,\bar u) \s_2(0) \big\rangle 
	= |1-u|^{4( h^\s_2 - h^\s_n)} g(u , \bar u) .
		\label{ssss2}
\ee
Channel (\ref{gx11}) gives an operator of dimension zero,  and channel (\ref{gx12}) an operator of dimension $\frac{2}{3} = h^\s_3$. 
This gives us the fusion rule
\be
[\s_n] \times [\s_n] = [{\mathds 1}] + [\s_3] + \cdots 
\ee
Of course, there are other twists in the r.h.s.~but they cannot be found from the four-point function we have began with, because of the condition (\ref{compto1}).
As discussed above, in channel (\ref{gx11}) the two twists $\s_2$ in the correlator have inverse cycles, hence it is necessary that the two twists $\s_n$ also be the inverse of each other; this gives $\mathds 1$ in the fusion rule. As for the channel (\ref{gx12}), we have seen that the cycles in $\s_2$ then only have one overlapping element, say, $\s_{(k\ell)} \s_{(k m)} = \s_{(k\ell m)}$. Hence for Eq.(\ref{compto1}) to be satisfied the two $\s_n$ operators must compose to $\s_n \s_n = \s_{(m\ell k)}$, which is why $\s_3$ appears.

\subsection{Non-BPS operators in the OPEs of $R^\pm_n$ with $\Oincov_2$} \label{SectOPEOintR}

We now turn to the limit $u \to 0$, where the interaction operator collides with either the Ramond field $R^+_n(0)$ or with the bare twisted field $\s_n(0)$, depending on the function we analyze, if either $G_R$ or $G_\s$. 
Now one can find all $2n$ solutions of Eq.(\ref{RamifEqHu}), viz.~$x = 0$ (with multiplicity $n-1$) and $x = -n$ (with multiplicity $n+1$), all contributing to the OPE limits.

\bigskip

The function $G_R(u,\bar u)$ gives the OPE $\Oint (u,\bar u) R^+_{[n]}(0)$. Using (\ref{xto0minnux}), 
\begin{align}
G_R(x^0_{\frak1}(u)) 
	&= u^{-\frac{5}{4} \frac{n-2}{n-1}} \big( C_{\frak1}  + c_{\frak1}  u^{1\over n-1} + \cdots \big) , \label{chanzero}		
\\
G_R (x^0_{\frak2}(u)) &= u^{-\frac{1}{4} \frac{3n-2}{n+1}} \big( C_{\frak2} +  c_{\frak2}  u^{1\over n+1}+ \cdots  \big) . \label{chanzero2}
\end{align}
Counting powers of $u$, we find that the OPE $\Oincov_2 R^+_n$ results in a twisted field  $Y^+_{m}$ which is positively R-charged with $j^3=\frac{1}{2}$, and has the holomorphic dimension
\be
h^Y_m = \frac{3}{2 m} + h^{\sigma}_{m} .
	\label{wheighY}
\ee
Channels $x^0_{\frak1}(u)$ and $x^0_{\frak2}(u)$ give $m = n-1$ and $m=n+1$, respectively.

The OPE $\Oint R^-_n$, is obtained in the limit $u \to 0$ of 
\be
\big\langle \Oint(\infty) R^-_{[n]} (u,\bar u ) \Oint (0) R^+_{[n]}(1) \big\rangle 
			=  |1 -u|^{4 - n} G_R ( u, \bar u) ,	
\ee
which follows from  the same procedure of fixing points used to find (\ref{mastROORuinf2}).
Since the factor of $(1-u)^{\frac{4-n}{2}}$ does not contribute to the leading term near $u = 0$, we immediately find the same expansion as before. Now the resulting fields 
$Y^-_{m}$ have the same dimensions (\ref{wheighY}), but opposite R-charge,  $j^3= - \frac12$.

In summary, we have found the fusion rules 
\be
[ \Oint ] \times [ R^\pm_{[n]} ]  = [ Y^\pm_{[n-1]} ] +  [  Y^\pm_{[n+1]} ] , \label {R-ope}
\ee
where the fields $Y^\pm_{m}$ have the dimension (\ref{wheighY}). 
The appearance of $m= n\pm1$ in the r.h.s.~is a basic consequence of permutation composition, see Eq.(\ref{fusionsimnn}).
We take the $Y^\pm_m$ to be normalized, so that (by charge conservation) the non-vanishing two-point functions are 
$$
\big\langle Y^\pm_m(z_1,\bar z_1) Y^\mp_m(z_2,\bar z_2) \big\rangle = |z_{12}|^{-4 h^Y_m} .
$$
Inserting the OPE back into the four-point function, the leading short-distance coefficients $C_{\frak a}$ give us information about the product of structure constants 
\bsub\label{StrConsOYR}\begin{align}
\big\langle R^-_n(\infty) \Oincov_2(1) Y^+_{n-1}(0) \big\rangle 
	\, 
\big\langle R^+_n(\infty) \Oincov_2(1) Y^-_{n-1} (0) \big\rangle &= | C_{\frak1}|^2 ,
\\
\big\langle R^-_n(\infty) \Oincov_2(1) Y^+_{n+1}(0) \big\rangle 
	\, 
\big\langle R^+_n(\infty) \Oincov_2(1) Y^-_{n+1}(0) \big\rangle &= | C_{\frak2}|^2 
\end{align}\esub
(Recall that we must take $|G_R(x^0_{\frak a}(u))|^2$.)
In the l.h.s.~we actually have products of conjugate three-point functions/structure constants, 
\be
 C^{R^\mp \Oincov Y^\pm}_{n,2,m} 
=
 \big\langle R^\mp_n \Oincov_2 Y^\pm_m \big\rangle
= 
 C^{R^\pm \Oincov Y^\mp}_{n,2,m}  ,
\ee
(with the twists in the subscripts)
and taking the explicit expressions for $C_{\frak a}$, found from the expansions (\ref{chanzero})-(\ref{chanzero2}), we get
\begin{align}
\log C^{R^\pm \Oincov Y^\mp}_{n,2,n-1}  &= - \frac{n+2}{2} \log(n-1) + \frac{n^2 - 4n - 2}{2(n-1)} \log n  ,
\\
\log C^{R^\pm \Oincov Y^\mp}_{n,2,n+1}  &= + \frac{n-2}{2} \log(n+1) - \frac{n^2 + 4n - 2}{2(n+1)} \log n .
\end{align}

\bigskip

A similar situation takes place in the OPE $\Oincov_2 \s_n$, found in the limit $u \to 0$ of $G_\s(u,\bar u)$. 
The two channels reveal twisted operators $\cal Y_{m}$ with zero R-charge and dimensions
\be
h^{\cal Y}_m = - \frac{5}{8} \left( \frac{m-2}{m} \right) + h^\s_{m} ,
\ee
in a fusion rule
\be
[ \Oint ] \times [ \s_{[n]} ]  = [ \cal Y_{[n-1]} ] +  [ \cal Y_{[n+1]} ] 
		\label {s-ope}
\ee
the terms in the r.h.s.~corresponding to $x^0_{\frak1}(u)$ and $x^0_{\frak2}(u)$, respectively.
The coefficients calculated from the expansion of $G_\s$ give structure constants as before:
\begin{align}
\log C^{\s \Oincov \cal Y}_{n,2,n-1}
		&= 	- \frac{(n+1)^2}{2n} \log (n-1)
			+ \frac{n^2 - 4n -1}{2(n-1)} \log n		
\\
\log C^{\s \Oincov \cal Y}_{n,2,n+1}
		&= 	+ \frac{(n-1)^2}{2n} \log (n+1)
			- \frac{n^2 + 4n -1}{2(n+1)} \log n		
\end{align}
where 
$\big\langle \s_n(\infty) \Oincov_2(1) \cal Y_{m}(0) \big\rangle \equiv C^{\s \Oincov \cal Y}_{n,2,m}$.

\bigskip

We have found that the operator algebras of Ramond fields with the deformation operator include non-BPS fields. 
These fields are consistent with the fractional spectral flow with $\xi =\frac{n}{n+1}$ of twisted non-BPS fields in the NS sector, recently  found \cite{deBeer:2019ioe} to be a part of  the OPEs of the deformation operator and NS chiral operators;
see the discussion in \S\ref{SectSpectralFlow}.
A complete study of the algebras found here requires knowledge of OPEs such as
$[ Y^\pm_{n\pm1}] \times [\Oint ]$.
For that,  the new fields have to be explicitly constructed. 
From our discussion of their properties, and in particular from the conformal dimension (\ref{wheighY}), we can infer that
\be
Y^{\pm}_{m} (0) = G^{\pm}_{-\frac{1}{2 m}} J^3_{-\frac{1}{m}} \sigma_{m}(0). \label{Ydesc}
\ee
This explicit construction should be sufficient for the study of the remaining OPEs by the computation of four-point functions such as 
$\langle Y^-_{[m]}(\infty) \Oint(1) \Oint(u,\bar u) Y^+_{[m]} \rangle$ (which, incidentally, can be computed with the same covering map used here).

\section{Analytic regularization and field renormalization} \label{SectRenormOfConf}

We now turn to the calculation of the conformal dimension of Ramond fields in the interacting SCFT$_2$. 
At second-order in perturbation theory, this requires computation of the integral (\ref{JintDef}), using the functions we have found in Sect.\ref{SectFourPointFuncts}.

\subsection{Dotsenko-Fateev integrals}	\label{SectDFIntegr}

We want to compute integrals  $J = \int d^2u \, G(u,\bar u)$,
given an analytic expression for 
\be
G(x) =C\, \frac{x^{\a_1} (x-1)^{\a_2} (x+n)^{\a_3} (x+n-1)^{\a_4}}{(x + \frac{n-1}{2})^{\a_5}}, 	
\quad
\text{with}
\quad
		\begin{cases}
		\a_1 + n - 2 = \a_4 - n
		\\
		\a_2 - n - 2 = \a_3 + n 
		\end{cases}
\label{GxApp}
\ee
from which $G(u,\bar u)$ is obtained by inversion of the map (\ref{ux}). 
Both the Ramond function (\ref{func}) and the bare twist function (\ref{funcsig}) have the form (\ref{GxApp}).
We can perform a change of variables from $u$ to $x$ in the integral which, taking the special relation between the exponents into account, becomes
\begin{align}	\label{Jintalphas}
J &=\int \! d^2 x \,|u'(x) G(x)|^2
	= (4nC)^2 \int\!d^2x \left| \frac{ \left[ x  (x+n-1)\right]^{\a_4-n} \left[ (x-1)(x+n) \right]^{\a_3+n} }{(x + \frac{n-1}{2})^{\a_5-2}} \right|^2 .
\end{align}
We then make the following change of variables \cite{Pakman:2009mi},
\be
y(x) = - \frac{4(x-1)(x+n)}{(n+1)^2} ,	\label{ChangofCorrd}
\ee
such that  every term in the new integrand is expressed simply in terms of $y$,
\begin{align}
& J(n) =  \frac{(4n C)^2}{4}
		\Big( 
			\frac{n+1}{2} 
		\Big)^{4 \left[ a+b+c + 1  \right] } 
		 \; I(n) , 	\label{hyper}
\\
& I(n) \equiv \int \! d^2y \;  |y|^{2a}|1-y|^{2b}|y-w_n|^{2c},  \label{DFitegrad}
\end{align}
where
\be
a = \a_3 + n , 
\quad
b = - \frac{\a_5-1}{2} ,
\quad
c = \a_4 - n ,
\quad
w_n = \frac{4n}{(n+1)^2} .		\label{DFexpnfora}
\ee
We will refer to $I(n)$ as a `Dotsenko-Fateev (DF) integral', as it has been studied in detail by Dotsenko and Fateev, as a representation of correlation functions in degenerate CFTs \cite{Dotsenko:1984nm,Dotsenko:1984ad,dotsenko1988lectures}.%
	\footnote{%
	Cf.~also Refs.\cite{Mussardo:1987eq,Mussardo:1987ab,Mussardo:1988av}.
	}

The properties of $I(n)$ crucially depend on the exponents of its critical points $y = \{0,w_n,1\}$. 
For example, the exponents for $G_R(x)$ are 
\be
a_R = \tfrac{1}{2} + \tfrac{1}{4}n , \qquad  b_R = - \tfrac{3}{2} ,  \qquad c_R = \tfrac{1}{2} - \tfrac{1}{4}n   
\label{abcwRamond}
\ee
thus, for general $n$, all three critical points are branching points. The integral diverges at $1$ and $\infty$, and vanishes at $0$ for all $n$;  at $w_n$, it converges for $n \leq 6$ and diverges for $n > 6$. 
Clearly, some regularization procedure is needed.
Following Ref.\cite{dotsenko1988lectures}, we now show that $I(n)$ can be expressed in terms of hypergeometric functions, leading to a regularization by analytic continuation.
We do this in two steps:

\begin{enumerate}

\item
Assume that the parameters $a,b,c$ are such that the DF integral exists. 

\item
Express the integrals in terms of an analytic function of $a,b,c$ that is well-defined also for values of $a,b,c$, such as (\ref{abcwRamond}), for which the original integral diverges.
(Such functions will turn out to be a product of hypergeometric and Gamma functions.) 
This leads to an extension of the definition of the integrals by their maximal analytic continuation.
\end{enumerate}
As we shall see, the procedure is consistent.

\begin{figure*} 
\centering
\includegraphics[scale=0.32]{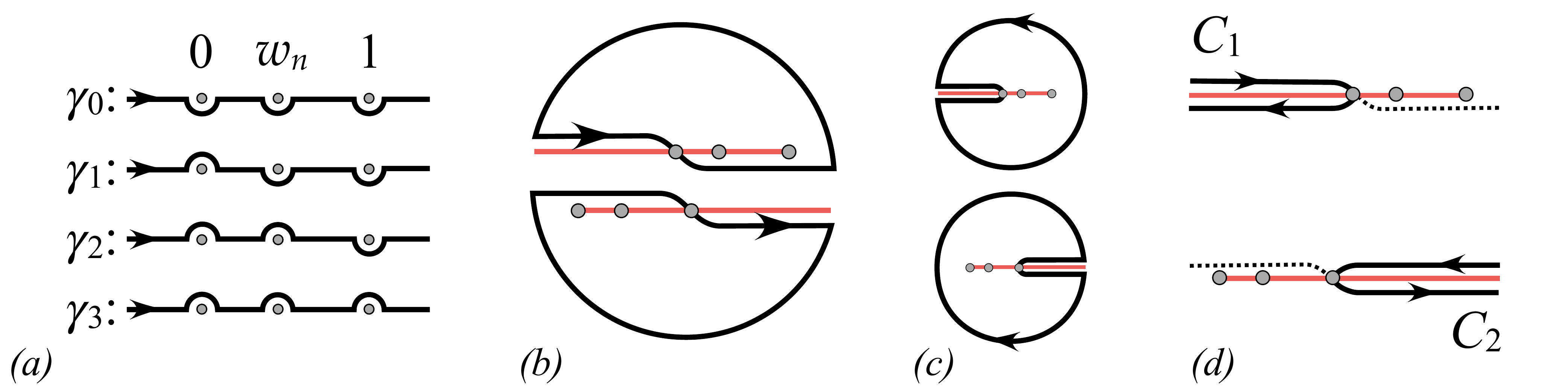}
\caption{
\textit{(a)} Contours for the Dotsenko-Fateev integral;
\textit{(b)} Closing $\gamma_1$ and $\gamma_2$;
\textit{(c)} Deformation;
\textit{(d)} Final contours (ignoring circles at infinity). Red lines indicate branching cuts.}
\label{ContoursAndDeformationLong}
\end{figure*}

Let us write $y \in \mathbb C$ as $y = y_1 + i y_2$ in (\ref{DFitegrad}),  and perform a rotation of $y_2$, such that $y_2  \mapsto i(1-2 i \vare) y_2$, with $\vare$ a positive arbitrarily small parameter. 
Defining $v_\pm = y_1 \pm y_2$ (where $y_2$ now refers to the new, rotated coordinate), and
expanding the integrand to first order in $\vare$, 
\begin{align*}
I	&= i \iint  dy_1  dy_2  \; \left[ y_1^2 - y_2^2 (1 -4i\vare)  \right]^a \big[ (y_1- 1)^2 - y_2^2 (1 - 4i\vare) \big]^b 
\\		
		&\hspace{2.7cm} \times \big[ (y_1 - w_n)^2 - y_2^2 (1 - 4i \vare) \big]^c
\\
	&=\frac{i}{2} \iint dv_-  dv_+  \; 
		\left[ \left\{ v_- -  i \vare (v_- - v_+) \right\}\left\{ v_+ + i \vare (v_- - v_+) \right\}  \right]^a 
\\	
	&\hspace{2.7cm} \times \left[ \left\{ v_- - 1 -  i \vare (v_- - v_+) \right\} \left\{ v_+ -1 + i \vare (v_- - v_+) \right\}  \right]^b 
\\		
		&\hspace{2.7cm} \times \left[ \left\{ v_- - w_n -  i \vare (v_- - v_+) \right\}\left\{ v_+ - w_n + i \vare (v_- - v_+) \right\}  \right]^c .
\end{align*}
The double integrals have been factorized into a product of two one-dimensional integrals,
\be
\begin{split}
I 	=\frac{i}{2}  \int_{-\infty}^{+\infty} \! dv_-   
		&\left[ v_- -  i \vare  (v_- - v_+)  \right]^a 
		 \left[ v_-  -1 -  i \vare  (v_- - v_+) \right]^b 
\\
	\times	&\left[ v_- - w_n -  i \vare  (v_- - v_+) \right]^c
\\		
	\quad \times \int_{-\infty}^{+\infty} \! dv_+  \; 
		&\left[ v_+ +  i \vare  (v_- - v_+) \right]^a 
		 \left[  v_+ -1 +  i \vare  (v_- - v_+) \right]^b 
\\
	\times &\left[ v_+ - w_n + i \vare (v_- - v_+) \right]^c
\end{split}
	\label{Idounbfact}
\ee
because the variable $v_\pm$ only appears in the $v_\mp$ integral multiplied by the infinitesimal parameter $\vare$. The effect of the $\vare$-terms is to specify how the otherwise real integrals of 
\be
f(\zeta) = \zeta^a (\zeta - 1)^b (\zeta - w_n)^c \ , \qquad \zeta \in {\mathbb C} \ , \qquad w_n \in (0 , 1) \subset {\mathbb R} ,
	\label{integrandfz}
\ee
go around the points $0,w_n, 1$.
To further disentangle the integrals, we split integration over $v_+$ at  $0,w_n,1$, so that $\vare$-terms can be ignored, while the $v_-$ integrals go around the contours $\gamma_k$ dictated by the infinitesimal terms $\vare (v_- - v_+)$ as in Fig.\ref{ContoursAndDeformationLong}\textit{(a)},
\be
\begin{split}
I = \frac{i}{2}  \Bigg[& \int_{-\infty}^0 dv_+ f(v_+) \int_{\gamma_0} dv_- f(v_-) +  \int_0^{w_n} dv_+ f(v_+) \int_{\gamma_1} dv_- f(v_-) 
\\
		+ &\int_{w_n}^1 dv_+ f(v_+)  \int_{\gamma_2} dv_- f(v_-)  + \int_1^\infty dv_+ f(v_+) \int_{\gamma_3} dv_- f(v_-)\Bigg] .
\end{split}
	\label{Isplitucontr}
\ee
For example, for $v_+ \in ( 0, w_n)$,
\[
\vare (v_- - v_+)|_{v_- = 0} > 0 , \qquad \vare (v_- - v_+) |_{v_- = w_n} < 0 , \qquad \vare (v_- - v_+)|_{v_- = 1} < 0 ,
\]
hence the contour $\gamma_1$ goes above $v_- = 0$, and below $v_- = w_n, 1$. 

%
%
%
%
%
%

The function $f(\zeta)$ has branch cuts, so closing the contours $\gamma_k$ with semi-circles is non-trivial.
Here is the point were our regularization procedure effectively starts. Assume that $a,b,c$ are such that the DF integral is convergent. Precisely, assume that 
\be
a > -1, \quad b> -1, \quad c > -1 , \quad 1+a+b+c < 0,	\label{assumpabc}
\ee
which ensures, respectively, convergence at the points $0,1,w_n, \infty$. 
Now close the contours $\gamma_k$ by making semicircles of radius $R \to \infty$ on the lower or upper plane. The curve $\ga_0$ passes below every branch point, hence close the contour with a clockwise semicircle $\Gamma^-$ on the lower plane; there are no poles inside $\gamma_0 \cup \Gamma^-$, and 
$ \lim_{R \to \infty} \int_{\Gamma^-} dv_- \ f(v_-) \to 0$ given our assumptions (\ref{assumpabc}); hence $\int_{\gamma_0} dv_- f(v_-) = 0$.
Similarly, $\int_{\gamma_3} dv_- f(v_-) = 0$, now with the contour on the upper plane.
Thus only two terms remain in Eq.(\ref{Isplitucontr}).

If we try to close $\ga_1$ or $\ga_2$ in Fig.\ref{ContoursAndDeformationLong}\textit{(a)}, we are deemed to cross branch cuts, and move to another Riemann sheet of $f(v_-)$. One way out of this is to cross the cut \emph{on a branching point}, where $f(v_-)$ is single-valued.
That the integral exists at the branching points is assured by our assumptions (\ref{assumpabc}).
Thus we choose the branch cuts to align with the Real axis in two different ways: for the integral over $\ga_1$ they extend to $-\infty$, and for $\ga_2$ they extend to $+\infty$; then we close the contours with semi-circles as in Fig.\ref{ContoursAndDeformationLong}\textit{(b)}.
In one case, we cross the real axis at $v_- = 0$, in the other at $v_- = w_n$.
Next, we deform the contours as in Fig.\ref{ContoursAndDeformationLong}\textit{(c)}. Given our assumptions (\ref{assumpabc}), as $R \to \infty$ the integral over the (almost closed) circle vanishes, and we have 
$\int_{\gamma_i} dv_-  f(v_-) = \int_{C_i} dv_- f(v_-)$
for $i = 1,2$, where the contours $C_i$ are shown in Fig.\ref{ContoursAndDeformationLong}\textit{(d)}.
Integration over $C_i$ is standard: the effect of coasting the two margins of a branch cut, turning at the branch point is to produce a phase $2i \sin(\pi \theta)$. 

Thus we arrive at the following form of (\ref{Isplitucontr}), 
\be
\begin{split}
I(a,b,c;w_n) = &- s(a) \tilde I_1(a,b,c;w_n) \  I_2(a,b,c;w_n) 
\\
	& - s(b) I_1(a,b,c;w_n) \ \tilde I_2(a,b,c;w_n) ,
\end{split}
	\label{DFmaster}
\ee
where $s(\theta) \equiv \sin(\pi \theta)$ and we have defined four `canonical integrals':
\bsub
\begin{align}
I_1(a,b,c;w_n) &\equiv \int_1^\infty dv_+ \, v_+^a (v_+ - 1)^b (v_+ - w_n)^c
\\
I_2(a,b,c;w_n) &\equiv \int_0^{w_n} dv_- \, v_-^a (1 - v_-)^b (w_n - v_-)^c
\\
\tilde I_1(a,b,c;w_n) &\equiv \int_{-\infty}^0 dv_+ \, (-v_+)^a (1 - v_+)^b (w_n - v_+)^c
\\
\tilde I_2(a,b,c;w_n) &\equiv  \int_{w_n}^1 dv_- \, v_-^a (1 - v_-)^b (v_- - w_n)^c
\end{align}
\esub
The $\tilde I_{1,2}$ can actually be written in terms of the $I_{1,2}$ with a different arrangement of their arguments: 
\be
\tilde I_1(a,b,c;w_n) =  I_1(b,a;c;1-w_n),	
\quad
\tilde I_2(a,b,c;w_n) = I_2(b,a;c;1-w_n).	
\label{tildeIfromI}
\ee
Also, by combining deformed contours such as the ones in Fig.\ref{ContoursAndDeformationLong}, it can be shown \cite{Dotsenko:1984nm} that $I_{1,2}(a,b,c,w_n)$ and $\tilde I_{1,2}(a,b,c,w_n)$, with the \emph{same} arguments, form a linear system:
\be
s(b+c) I_1 = s(a) \tilde I_1 - s(c) \tilde I_2 \,,
\quad
s(b+c) I_2 = - s(a+b+c) \tilde I_1 - s(b) \tilde I_2 .%
\label{SystemDF}
\ee

The four canonical integrals are proportional to the Euler representation of the hypergeometric function \cite{NIST:DLMF151}, 
\begin{flalign}
\begin{split}
& \int_0^1\! dt \, t^{\b -1} (1 - w t)^{-\a} (1 - t)^{\ga - \b - 1}
	= \frac{\Gamma(\b) \Gamma(\ga - \b)}{\Gamma(\ga)} F(\a,\b;\ga;w)  
\\		
&\text{valid for} \quad |\rm{arg}(1-w)| < \pi , \quad 0 < \rm{Re}(\b) < \rm{Re}(\ga) . \label{Restabc1}
\end{split}
\end{flalign}
With the substitution $t = 1 / v_+$ in $I_1$, and $t = v_- / w_n$ in $I_2$, we find 
\begin{align}
I_1(a,b;c;w_n) &= \frac{\Gamma(-1-a-b-c) \Gamma(1+b)}{\Gamma(-a-c)} F(-c, -1-a-b-c ; -a-c ; w_n) ,
	\label{IHyperabc1}
\\
I_2(a,b,c;w_n) &=  \frac{\Gamma(1+a) \Gamma(1+c)}{\Gamma(2+a+c)} w_n^{1+a+c} F(-b, 1+a ; 2+a+c ; w_n) .
		\label{IHyperabc2}
\end{align}
The restrictions (\ref{Restabc1}), required for both integrals to be represented by hypergeometrics, translate to $a,b,c$ as
\be
-1 < a < 1 , \quad -1 < b < -a , \quad -1 < c < -1-a-b  ,	\label{DomaValid}
\ee
and also $0 < w_n < 1$, cf.~(\ref{DFexpnfora}).
These conditions are consistent with our starting hypothesis (\ref{assumpabc}), therefore Eq.(\ref{DFmaster}) can be read as a product of hypergeometric and Gamma functions.

The `canonical functions' (\ref{IHyperabc1}) and (\ref{IHyperabc2}) are analytic functions of each of the parameters $a,b,c$, on the domain of validity (\ref{DomaValid}). 
This is evident for the Gamma functions, and is also true for the hypergeometrics, see \cite[\S2.1.6]{bateman1953higher}.
Note that in (\ref{IHyperabc1}) and (\ref{IHyperabc2}) what actually appears is the `\emph{regularized} hypergeometric function'
 \be
 {\bf F}(\a,\b;\ga;w) \equiv  \frac{1}{ \Gamma(\ga)} F(\a,\b ;\ga; w) ,
 \ee
 which is an entire function of $\a,\b,\ga$  \cite[\S2.1.6]{bateman1953higher}.
 In particular, ${\bf F}(\a,\b;\ga;w)$ is regular at $\ga = - k$, with $k \in \mathbb N$, where the Gamma function develops a pole and \cite[\S15.2]{NIST:DLMF}
\be
{\bf F}(\a,\b ;-k; w)  = \frac{\Gamma(\a+k+1) \Gamma(\b +k+1)}{\Gamma(\a) \Gamma(\b)(k+1)!} w^{k+1} F(\a+k+1, \b+k+1; k+2 ; w) .
\label{RegHyperInte}
\ee
Hence $I(a,b,c;w_n)$ is analytic in $a,b,c$ separately. 
Consequently, an analytic continuation of $I(a,b,c;w_n)$ to outside of the domain of definition (\ref{DomaValid}) is \emph{unique}, when it exists. We take this analytic continuation to be the definition of the DF integral (\ref{DFitegrad}) for arbitrary parameters. 
Note that it is not precluded that, outside the domain (\ref{DomaValid}), $I(a,b,c;w_n)$ might develop a singularity --- there may be a barrier to the analytic continuation --- it just happens that, for the applications below, the continuation is, indeed, (almost) always well-defined.

\subsection{The integral for R-charged Ramond fields}	\label{SecRenormCharRam}

Let us apply our results to the Ramond function (\ref{func}).
As noted before, the parameters (\ref{abcwRamond}) \emph{do not} lie within the domain (\ref{DomaValid}), hence we are indeed using the analytic continuation.
Eqs.(\ref{IHyperabc1}), (\ref{IHyperabc2}), (\ref{tildeIfromI}) yield
\bsub\begin{align}
I_1(a_R,b_R,c_R) &= \frac{\pi (4-n^2)}{ 32} \, w^2 \, F ( \tfrac{3}{2}, \tfrac{3}{2} + \tfrac{1}{4} n ; 3 ; w_n)		\label{hyperDFI1}
\\
I_2(a_R,b_R,c_R)	&= \frac{1}{s(\tfrac{1}{2} - \tfrac{n}{4}) } I_1	(a_R,b_R,c_R)	\label{hyperDFI2}
\\
\tilde I_1 (a_R,b_R,c_R) &= - \frac{2 \sqrt\pi \Gamma(\tfrac{3}{2} + \tfrac{n}{4})}{ \Gamma(1 + \tfrac{n}{4})}F(- \tfrac{1}{2} + \tfrac{n}{4} , - \tfrac{1}{2} ; 1 + \tfrac{n}{4} ; 1-w_n )	\label{tildI2odda}
\\
\tilde I_2(a_R,b_R,c_R) &= - \frac{2 \sqrt\pi \Gamma( \tfrac{3}{2} - \tfrac{n}{4} )}{ \Gamma( 1 - \tfrac{n}{4})} (1-w)^{-n/4} F(- \tfrac{1}{2} - \tfrac{n}{4} , - \tfrac{1}{2} ; 1 - \tfrac{n}{4} ; 1-w_n )	\label{tildI2odd}
\end{align}\esub
Several observations are in order.
The expression (\ref{hyperDFI1}) does not correspond immediately to the formula (\ref{IHyperabc1}), because here we have $\Gamma(-1)$ in the denominator. In this case, we must use Eq.(\ref{RegHyperInte}) to find  the correct expression for $I_1$ in (\ref{hyperDFI1}). Expression (\ref{hyperDFI2}) can be found immediately from (\ref{IHyperabc2}). The factor $s(c)$ in (\ref{hyperDFI2}) can be found either from $\Gamma(z) \Gamma(1-z) = \pi / \sin (\pi z)$, or from the linear system (\ref{SystemDF}), by noting that in the present case we have
\be
s(a_R+c_R) = 0 ,
\qquad
s(a_R) = s(c_R) ,
\qquad
s(a_R+b_R+c_R) = - s(b_R) = -1  .	\label{sabcRam}
\ee
Eqs.(\ref{tildI2odda}) and (\ref{tildI2odd}) follow immediately from (\ref{tildeIfromI}), but
(\ref{tildI2odd}) is  only valid when $n$ is odd. For $n$ even there are two cases.
When $n = 4(k+1)$ a pole of the Gamma function in the denominator of (\ref{tildI2odd}) requires that we use Eq.(\ref{RegHyperInte}) again, leading to 
$\tilde I_2 = \tilde I_1.$ 
This can also be found from the linear system (\ref{SystemDF}) by noting that, besides (\ref{sabcRam}), now $s(b_R+c_R) = 0$.

All of the peculiarities above are taken into account if we simply replace the hypergeometrics by the well-behaved regularized hypergeometric, 
\bsub\begin{align}
I_1(a_R,b_R,c_R) &= \frac{\pi (4-n^2)}{ 32} \, w_n^2 \, F ( \tfrac{3}{2},  \tfrac{6+n}{4}  ; 3 ; w_n)
\\
I_2 (a_R,b_R,c_R) &= \tfrac{1}{2}  \Gamma ( \tfrac{6 - n}{4} ) \Gamma (\tfrac{6+n}{4} ) w_n^2 F ( \tfrac{3}{2},  \tfrac{6+n}{4}  ; 3 ; w_n)
\\
\tilde I_1(a_R,b_R,c_R) &= - 2 \sqrt\pi \, \Gamma( \tfrac{6+n}{4})\,  {\bf F}( \tfrac{n-2}{4} , - \tfrac{1}{2} ;  \tfrac{n+4}{4} ; 1 - w_n )
\\
\tilde I_2(a_R,b_R,c_R) &= - \frac{2 \sqrt\pi}{(1-w_n)^\frac{n}{4}} \, \Gamma(  \tfrac{6-n}{4} )  \,  {\bf F}(- \tfrac{n+2}{4} , - \tfrac{1}{2} ;  \tfrac{4-n}{4} ; 1-w_n )	\label{tildI2odd}
\end{align}\label{CanIinteHyper}\esub
We can now use Eqs.(\ref{DFmaster}) and (\ref{hyper}) to write
\be	\label{Jrncosetc}
J_R(n) = - \left( \frac{n+1}{32 n} \right)^2 \Big[ \cos \left(\frac{n \pi}{4}\right) \tilde I_1(n) I_2(n) + I_1(n) \tilde I_2(n) \Big] , 
\quad
	\begin{cases}
	n \neq 4k + 2 
	\\
	k \in \mathbb N
	\end{cases}
\ee
Before we analyze this result further, let us consider what happens if $n = 4k+2$.


\bigskip

\noindent
{\bfseries The case $n = 4k +2$}

\vspace{1mm}

\noindent
When $n = 4k +2$, a pole of the Gamma function appears in the \emph{numerator} of (\ref{tildI2odd}), so $I(n)$ is infinite. 
We can isolate the divergence, however. First, we list again the four canonical integrals, now in terms of $k = \frac{n-2}{4}$,
\bsub\begin{align}
I_1(k) &= -\frac{32 \pi  k (k+1) (2 k+1)^2}{(4 k+3)^4} \, F \left(\tfrac{3}{2}, k+2 ; 3 ; \tfrac{8 (2 k+1)}{(4 k+3)^2}\right)
	\label{RamonDivhyperDFI1}
\\
I_2(k)	&=\frac{32 (2 k+1)^2}{(4 k+3)^4} \, \Gamma (1-k) \Gamma (k+2) \,  F \left( \tfrac{3}{2} , k+2 ; 3 ; \tfrac{8 (2 k+1)}{(4 k+3)^2}\right) 
	\label{RamonDivhyperDFI2}
\\
\tilde I_1(k) &= -2 \sqrt{\pi } \, \Gamma (k+2) \, {\bf F} \left(-\tfrac{1}{2} , k ; k+ \tfrac{3}{2} ; \tfrac{(4 k+1)^2}{(4 k+3)^2}\right) 
	\label{RamonDivhyperDFI1t}
\\
\tilde I_2(k) &= - \frac{2 \sqrt{\pi } (4 k+3)^{2 k+1}}{(4 k+1)^{2 k+1}}  \, \Gamma (1-k) \, {\bf F} \left(-\tfrac{1}{2} , -k-1 ;\tfrac{1}{2} - k ; \tfrac{(4 k+1)^2}{(4 k+3)^2}\right) 
	\label{RamonDivhyperDFI2t}
\end{align}\label{RamDivIsall}\esub
Here we note that in this case we have $s(a_R) = s(c_R) = 0$ besides (\ref{sabcRam}), and the linear system (\ref{SystemDF}) is not valid anymore. This is related to the fact that there is now only one branch point in the canonical integrals, instead of the three branchings of the general case. 
Eq.(\ref{DFmaster}) is, however, still valid.
Moreover, we have $I_2(k) = - I_1(k) / \sin(\pi k)$. The sine is cancelled in Eq.(\ref{DFmaster}), 
\be
I(k) = - I_1(k)  \big( \tilde I_1(k)  +   \tilde I_2(k)	\big) .	\label{I1tiI1I2Tid}
\ee
Now the only singularity comes from the pole of $\Gamma(1-k)$ in (\ref{RamonDivhyperDFI2t}). Making $k \to k +\epsilon$, we have (cf. \cite{bateman1953higher}, \S1.17 Eq.(11))
\be
\Gamma(1-k-\epsilon) = \frac{(-1)^{k-1}}{(k-1)!} \left[ - \frac{1}{\epsilon} + \psi(k) + {\rm O}(\epsilon) \right] ,	\label{Gammaepspsi}
\ee
from which we separate the finite part and the divergence:
\begin{align}
\tilde I_2(k) &\equiv \frac{1}{\epsilon} \tilde I_2^{sing}(k) + \tilde I_2^{reg}(k)  
\\
\tilde I_2^{sing}(k) &= \frac{(-1)^{k} 2 \sqrt\pi (4 k+3)^{2 k+1} }{(4 k+1)^{2 k+1} (k-1)!}  \,  {\bf F} \left(-\tfrac{1}{2} , -k-1 ;\tfrac{1}{2} - k ; \tfrac{(4 k+1)^2}{(4 k+3)^2}\right) 
\\
\tilde I_2^{reg}(k)  &= \frac{(-1)^{k-1} 2 \sqrt\pi (4 k+3)^{2 k+1} \psi(k) }{(4 k+1)^{2 k+1} (k-1)!}  \,  {\bf F} \left(-\tfrac{1}{2} , -k-1 ;\tfrac{1}{2} - k ; \tfrac{(4 k+1)^2}{(4 k+3)^2}\right) 
\end{align}
where $\psi(\zeta)$ is the digamma function.
Using (\ref{I1tiI1I2Tid}) and (\ref{hyper}), we end up with 
\begin{align}
J_R(k) &= J_R^{reg}(k) + \e^{-1} J_R^{sing}(k)
\\
J_R^{reg}(k) &= - \left( \frac{3 + 4k}{128(1 + 2k)} \right)^2 \ I_1(k) \ \big[ \tilde I_1(k)  +   \tilde I_2^{reg} (k)	\big] 	, \label{FineiJRk}
\\
J_R^{sing}(k) &= -  \left( \frac{3 + 4k}{128(1 + 2k)} \right)^2 \ I_1(k) \tilde I_2^{sing} (k) .
	\label{JRsingk}
\end{align}
which is the final regularized expression for $J_R$ when $n = 4k+2$.



%
%
%
\begin{figure}[t] 
\centering
\includegraphics[scale=0.7]{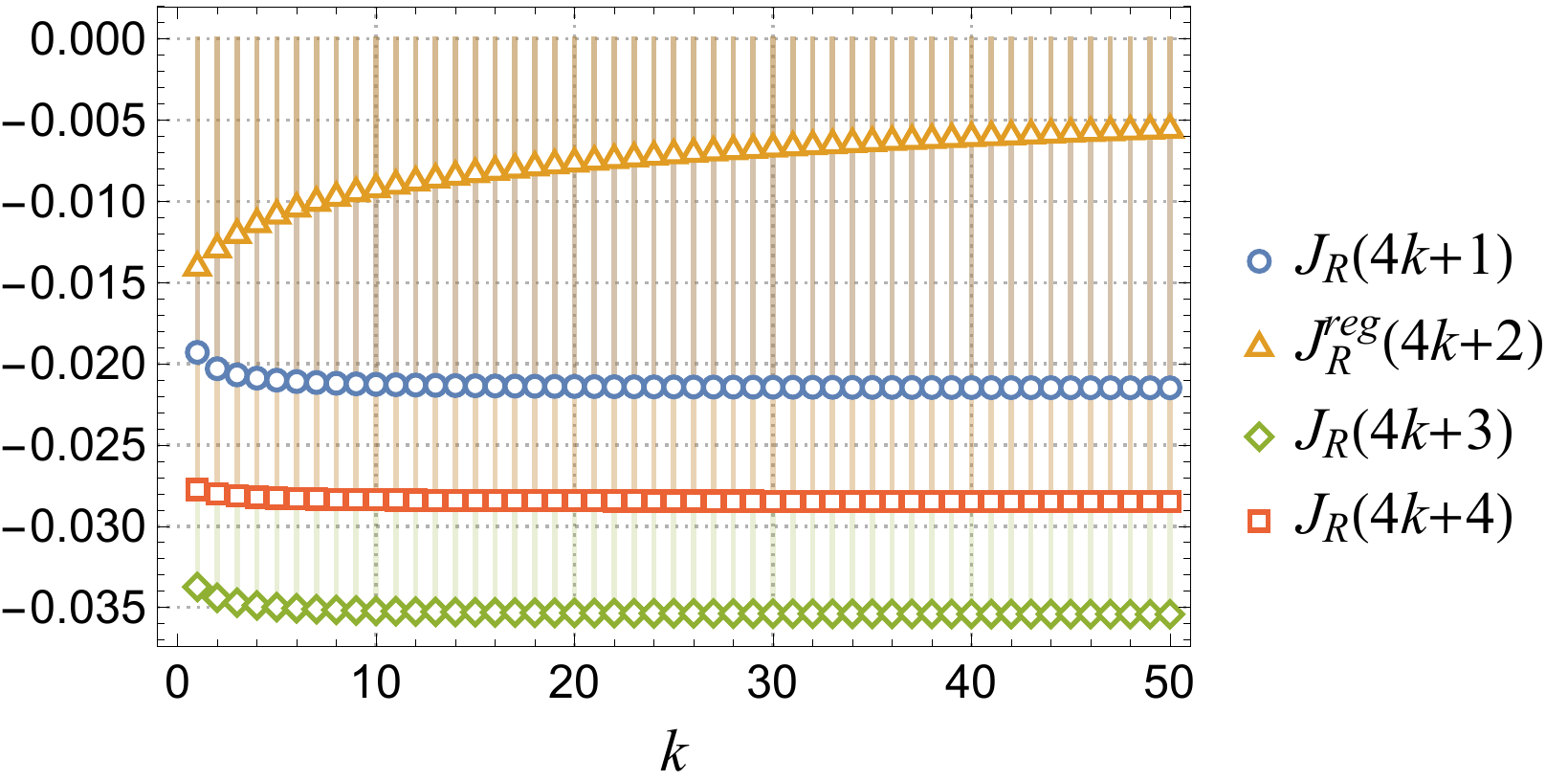}
\caption{$J_R(n)$ (and its regularization) for charged Ramond fields.}
\label{UnifiedRamondPlot}
\end{figure} 
%
%
%


\bigskip

\noindent
{\bfseries Comments}

\vspace{1mm}

\noindent
We present a unified plot of $J_R(n)$ for every $n$ in Fig.\ref{UnifiedRamondPlot};
for $n = 4k+2$, we plot the regularized function $J_R^{reg}(k)$.
One can distinguish a peculiar ``almost periodicity'' of the function, with period 4. We believe that this might be related to some combinatoric relation between the twists of $R^\pm_n$ and the twists of the interaction operators appearing in the four-point function.
As can be seen from Fig.\ref{UnifiedRamondPlot}, $J_R(n)$ stabilizes around small, negative values for large $n$. As a reference, for $k =30$ we have
\begin{alignat*}{3}
&
J_R(4k+1) \approx - 0.0215279
\qquad&&
J_R^{reg}(4k + 2) \approx - 0.0036010 
\\
&
J_R(4k+3) \approx - 0.0354618 
\qquad&&
J_R(4k + 4) \approx - 0.0284950  
\end{alignat*}
Note that an analytic form of $J_R(n)$ for large $n$ is very hard to find because it involves taking simultaneous limits of the multiple arguments of the hypergeometric function.

\subsection{The integral for bare twists}

 The function $G_\s(x)$ also has the form (\ref{GxApp}), and
\begin{align}
J_\s (n) = \int  d^2x \; |u'(x)G_\s(x)|^2  =  \left[ \tfrac{1}{2} n(n+1)C_\s \right]^2 \; I(n) , \label{Jsigma}
\end{align}
where $I(n)$ is a DF integral (\ref{DFitegrad}) with exponents
\be
a_\s = \frac{(n+1)^2}{4n} , \qquad  b_\s = - \frac{3}{2} ,  \qquad c_\s = - \frac{(n-1)^2}{4n}  .
\label{abcwBare}
\ee
The canonical integrals 
\bsub\begin{align}
I_1(a_\s,b_\s,c_\s) &= -\frac{\pi  (n-1)^2}{2 (n+1)^2}  \, F \left( \tfrac{3}{2} , \tfrac{1+6n +n^2}{4n} ; 3 ; w_n \right)			\label{twisthyperDFI1}
\\
I_2(a_\s,b_\s,c_\s)	&= \frac{8 n^2}{(n+1)^4} 
			\Gamma \left( 1 - \tfrac{(n-1)^2}{4n} \right) 
			\Gamma \left( 1+\tfrac{(n+1)^2}{4n} \right)
			F \left( \tfrac{3}{2} , \tfrac{1+6n+n^2}{4n} ; 3; w_n \right)
	\label{twisthyperDFI2}
\\
\tilde I_1(a_\s,b_\s,c_\s) &= -2 \sqrt{\pi } \, 
			\Gamma \big( 1 + \tfrac{(n+1)^2}{4 n} \big) \, 
			{\bf F} \left(- \tfrac{1}{2} , \tfrac{(n-1)^2}{4 n} ;  \tfrac{1+ 4n + n^2}{4n} ; 1 - w_n \right)
	\label{twisthyperDFI1t}
\\
\tilde I_2(a_\s,b_\s,c_\s) &= -2 \sqrt{\pi } \,  
			\Gamma \big( 1 - \tfrac{(n-1)^2}{4 n} \big)  \, 
			(1 - w_n)^{-\frac{1+n^2}{4 n}} \, 
			{\bf F} \left(-\tfrac{1}{2}, - \tfrac{(1+n)^2}{4 n} ; - \tfrac{1 - 4n + n^2}{4n} ; 1 - w_n \right)
\label{twisthyperDFI2t}
\end{align}\label{twisIsall}\esub
are all well-defined and convergent for all values of $n\in \mathbb N$ --- 
the arguments of the Gamma functions are never a negative integer (for $n>1$).

\begin{figure}[t] 
\centering
\includegraphics[scale=0.7]{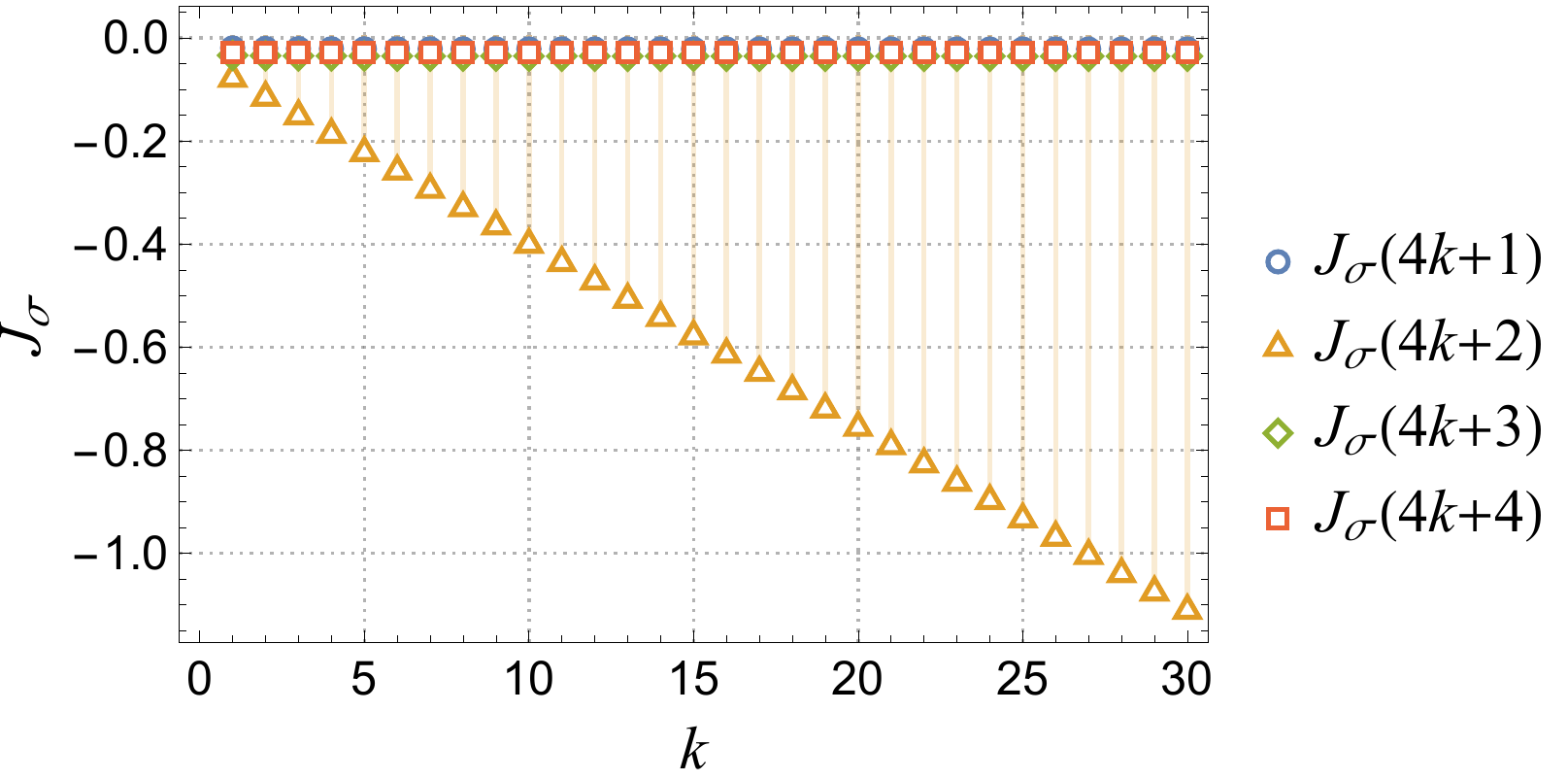}
\caption{Numerical result of the integral (\ref{Jsigma}) for twist fields.}
\label{PlotIntegralsTwits}
\end{figure}

We plot the values of 
\be
J_\s(n) = - \left( \frac{n+1}{32 n} \right)^2 \left[  s (a) \tilde I_1(n) I_2(n) + s(b) I_1(n) \tilde I_2(n) \right]
\ee
in Fig.\ref{PlotIntegralsTwits}.
We can see again an approximate periodicity, with period 4, similar to what happens in the Ramond case. 
We can give the following numerical values for $k = 30$:
\begin{alignat*}{3}
&
J_\s(4k+1) \approx -0.0214398 
\qquad&&
J_\s(4k + 2) \approx - 1.11106  
\\
&
J_\s(4k+3) \approx - 0.035381 
\qquad&&
J_\s(4k + 4) \approx - 0.028456 
\end{alignat*}
to be compared with the corresponding values for $J_R(n)$ given above.
For $n = 4k+2$, $J_\sigma(n)$ grows with $n$, instead of stabilizing around a small value; note that these values of $n$ are also those for which the Ramond integral $J_R(n)$ diverged, and had to be regularized.

%
%
%
%
%
%

\subsection{Renormalization of Ramond and  twist fields}

The renormalized dimension of the Ramond operators is given by Eq.(\ref{DeltccorinG}).
To first order in $\la$, it would be proportional do the structure constant of the three-point function 
\be
\langle R^-_{[n]}(\infty) \Oint(z )R^+_{[n]} (0)\rangle = 0 ,
\ee 
which vanishes at the free orbifold point because  $R^-_{[n]}$ does not appear in the OPE $\Oint(z, \bar z) R^+_{[n]} (0)$. 
We are thus left with a correction only at order $\la^2$, 
\be	\label{DeltaRnlaJR}
\Delta^R_n (\la) = \frac{n}{2} + \frac{\pi}{2}  \la^2 | J_R(n) | + \cdots,
\ee
where $\frac{1}{2}n = h^R_n(0) + \tilde h^R_n(0) = \Delta^R_n(0)$ is the (total) dimension of $R^\pm_{[n]}(z,\bar z)$ in the free theory.
Here it should be understood that $J_R(4k + 2) \equiv J_R^{reg}(k)$. The renormalized fields are therefore given by
\be
R^{\pm(ren)}_{[n]}(z,\bar z) = \La^{\frac{\pi}{2} \la^2  J_R(n)} R^\pm_{[n]}(z,\bar z) ,
\ee
where $\La \ll 1$ is the cutoff appearing in (\ref{PErtuTwoPtOOsecla}).
We can also give the renormalization of the structure constant; from Eq.(\ref{deltaCstrc}), 
\be
\big\langle R^-_{n}(\infty) \Oint(1) R^+_{n}(0) \big\rangle = \la J_R(n) + \cdots
\ee

We thus have concluded that for generic $n < N$ the dimension of the Ramond field slightly increases in the perturbed theory.
The case $n = 2$, which is the smallest value of $n$ for a twisted field,%
	\footnote{%
	We note that the  results above do not apply directly to the untwisted Ramond field ($n = 1$). In particular, the DF integral has a different structure, with one less critical point.}
is special. 
The function (\ref{func}) 
loses the singularity at $x = 0$. It is interesting to see how this arises as a consequence of the permutation structure of the twists: we have seen that the OPE channel with $x \to 0$ results in a field with twist $n -1$; in this case that would be an untwisted field. As a consequence of Eq.(\ref{compto1}), this would mean that the 2-cycles of $\Oincov_2$ and $R^\pm_2$ entering the correlators in this channel are all the same; but such a correlator would involve only ${\bf s} = 2 = n$ copies instead of $n+1$, and therefore require a genus-one covering surface.
Most importantly,  \emph{the $n=2$ twisted Ramond fields do not renormalize} --- it can be easily checked that 
\be
J_R(2) = 0 .
\ee
For that, it suffices to look at Eqs.(\ref{Jrncosetc}) and (\ref{CanIinteHyper}), since 
$\cos (\frac{\pi}{2}) = 0$ and $I_1(2) = 0$ while the other canonical functions are finite.

\bigskip

The same analysis holds for the bare twists: their dimension in the perturbed theory becomes
\be
\Delta^\s_n (\la) = \frac{1}{2}\left(n - \frac{1}{n} \right) + \frac{\pi}{2} \la^2 |J_\s(n)| + \cdots,
\ee
and the renormalized twist operators are 
\be
\s_{[n]}^{(ren)}(z,\bar z) = \La^{\frac{\pi}{2} \la^2  J_\s(n)} \s_{[n]} (z,\bar z) .
\ee
The structure constant 
\be
\langle \s_{[n]}(\infty) \Oint(1) \s_{[n]}(0) \big\rangle = \la J_\s(n) + \cdots
\ee
which also vanishes in the free theory, acquires a non-vanishing value at first-order.

\bigskip

%

The regularization of the divergent integral $J$ described in this section  gives well-defined, finite two-point functions in the deformed theory, to second order in $\la$.
Here we have considered the renormalization of bare twists and Ramond fields, but the method is more general, and can be applied to all sectors of the SCFT$_2$. 
Our procedure relied on the fact that $J$ can be reduced to a Dotsenko-Fateev integral for the functions $G_R(x)$ and $G_\s(x)$. This, in turn, relied on the structure of these functions, which had the form (\ref{GxApp}). It is not hard to check that,
for \emph{any primary twisted}  $\scr O_{[n]}$ that we insert in the general correlation function (\ref{4pointR1Gu}),
the corresponding $G(x)$ \emph{always} has the form (\ref{GxApp}), including the specific relations between the pairs of exponents $\a_1,\a_3$ and $\a_2,\a_4$.

To see that this is true, one can reverse-engineer the reasoning developed in Sect.\ref{SectOPElimits}. 
Given an operator $\scr O_n$ consider the correlator
$$
G(u, \bar u) = \big\langle \scr O_{[n]}^\dagger(\infty) \Oint(1) \Oint (u , \bar u) \scr O_{[n]}(0) \big\rangle 	.
$$
This function must be singular in the short-distance limit $u\to 0$, and consistent with the OPE rule (\ref{OPE}). 
The associated function $G(x)$ must therefore be singular when $x$ goes to  one of the values $x^0_{\frak a}$ for $u\to 0$, or $x^\infty_{\frak a}$ for $u \to \infty$, where $x^0_{\frak a}$, $x^\infty_{\frak a}$ are the channels in the limits $u\to 0$ and $u \to \infty$, respectively; see App.\ref{AppAsympOPE}. This fixes the numerator of 
\be
G(x) = C \frac{
		(x - x^0_{\frak1})^{\a_1}
		(x - x^\infty_{\frak2})^{\a_2}
		(x - x^0_{\frak2})^{\a_3}
		(x - x^\infty_{\frak1})^{\a_4}
		}{
		(x - x^1_{\frak2})^{\a_5}
		} 
	\label{Gxfinal}	
\ee
while the denominator is fixed similarly by the channels in $u\to 1$.
But (\ref{Gxfinal}) is just another way to write (\ref{GxApp}).
Note that this argument only makes use of the properties of the function $u(x)$ and its inverses, i.e.~only on the structure of the \emph{twists} in the correlator --- \emph{not} on the specifics of $\scr O_n$ nor, even, on the properties of $\Oincov_2$. Thus $\scr O_{[n]}$ can be, say, a primary NS field, an R-charged or R-neutral Ramond ground state, or a bare twist field; also, we can replace $\Oint$ by, say, the simplest chiral NS primaries $O^{(p,q)}_{[2]}$ (defined e.g.~in \cite{Skenderis:2008qn}).

Having proved that $G(x)$ must have the structure (\ref{Gxfinal}),
it remains for us to show that the exponents satisfy the two relations in (\ref{GxApp}). This is also a consequence of the OPEs. Take the channel $x^0_{\frak1}$ in the limit $u \to 0$. We have the OPE $\Oincov_2 \scr O_n \sim \scr X_{n-1}$ for some operator of twist $n-1$, whose dimension  is fixed by the power of $u$ appearing in $G(x^0_{\frak1}(u))$. 
Since $G(x^0_{\frak1}(u)) \sim [x^0_{\frak1}(u) ]^{\a_1}$, using (\ref{xto0minnux}) we have $G(x^0_{\frak1}(u)) \sim u^{\frac{\a_1}{n-1}}$, hence the holomorphic dimension of $\scr X_{n-1}$ is
\be
h^{\scr X}_{n-1} = \frac{\a_1}{n-1} + h^{\scr O}_n + h^{\Oincov}_2 .	\label{dimscrX}
\ee
Now, in the limit $u \to \infty$, we will have the OPE 
$\Oincov_2 \scr O_n^\dagger \sim \scr X_{n-1}^\dagger$. Using (\ref{xto0minnuxInf}), we now have 
$G( x^\infty_{\frak1}(u)) 
					 \sim u^{\frac{\a_4}{n-1}}$,
and the dimension of $\scr X^\dagger_{n-1}$ is
\be
h^{\scr X^\dagger}_{n-1} = \frac{\a_4}{n-1} + h^{\scr O}_n - h^{\Oincov}_2 ,
	\label{dimscrXdag}
\ee
with a minus sign in front of $h^{\Oincov}_2$ because we must conjugate $\Oincov_2$ to $\infty$.
But $\scr X_m$ and $\scr X^\dagger_m$ have the same dimension, so subtracting Eqs.(\ref{dimscrX}) and (\ref{dimscrXdag}) we find
$$
\frac{\a_1 - \a_4}{n-1} = 2 h^{\Oincov}_2 
$$
which, since $h^{\Oincov}_2 = 1$, gives the first relation in (\ref{GxApp}). The second relation, between $\a_1$ and $\a_3$, is found similarly in the channels $x^0_{\frak2}$ and $x^\infty_{\frak2}$, completing the proof that (\ref{GxApp}) holds in general.

Thus we have shown that, for any primary twisted filed  $\scr O_{[n]}$, we can always reduce $J_{\scr O}$ to a Dotsenko-Fateev integral, for some set of parameters $a,b,c$. Then, we can apply our regularization procedure and subsequent renormalization of the two-point function $\langle \scr O_{[n]}^\dagger \scr O_{[n]} \rangle$ --- \emph{if} that is necessary.
A very important example of fields for which there is \emph{no} renormalization is the class of BPS-protected NS chiral twisted fields.
Explicit computation of their non-renormalization was given in \cite{Pakman:2009mi},
for $O^{(0,0)}_n$, the lowest-weight operator in the $n$-twisted sector of the NS chiral ring \cite{Jevicki:1998bm}, with
$h^{\rm{NS}}_n = \frac{n-1}{2} = j^3$.
(The descendants of $O^{(0,0)}_2$ give the deformation operator $\Oincov_2$.)
The four-point function $G_O(x)$ was found by the same method of Sect.\ref{SectFourPointFuncts}, see Eq.(D.6) of Ref.\cite{Pakman:2009mi}.
It has the form (\ref{GxApp}), and gives rise to a Dotsenko-Fateev integral with exponents
$a_O = 1$,  $b_O = - \tfrac{3}{2}$, $c_O = 0$.
Since $c_O = 0$, the integral (\ref{DFitegrad}) simplifies, and can be computed directly in terms of Gamma functions, as done in App.D of Ref.\cite{Pakman:2009mi}, without the need to resorting to the hypergeometric regularization machinery.
Nevertheless, it is interesting to confirm that our formulae do give the same result, i.e.~$J_O = 0$.
Inserting $a_O$, $b_O$ and $c_O$ into our canonical functions (\ref{IHyperabc1})-(\ref{IHyperabc2}), we find
$$
I_1 = 0 ,
\quad
 I_2 = -4 - \frac{2 \left(w_n-2\right)}{\sqrt{1-w_n}} ,
\quad
\tilde I_1= -4 ,
\quad
\tilde I_2 = \frac{2 (w_n-2)}{\sqrt{1-w_n}}
$$
and $s(a_O) = 0$, so
\be	\label{JOeq0}
J_O(n) = - \left[ \tfrac{1}{2}  n (n+1) C_O \right]^2 \Big( s ( a_O ) \tilde I_1(n) I_2(n) +  s( b_O ) I_1(n) \tilde I_2(n) \Big) = 0
\ee
as expected.

Let us point out  that our regularization and renormalization procedure is even \emph{more} general. It can be extended almost intactly for the analysis of two-point functions of operators with a more complicated twist structure. In Ref.\cite{Lima:2020nnx} we have studied the double-cycle composite Ramond fields $R^\pm_{[n]}R^\pm_{[m]}(z,\bar z)$. In this case, the covering map is more complicated, and, correspondingly, so is the form of $G(x)$ which generalizes (\ref{GxApp}); but just as explained above, there are relations between exponents which allow a transformation of $J(n,m)$ into a Dotsenko-Fateev integral, and then everything follows as in here.


\subsection{On spectral flow}	\label{SectSpectralFlow}

The spectral flow automorphism of the $\cal N = 4$ super-algebra \cite{Schwimmer:1986mf} acts on the Virasoro and R-current modes as
\begin{align}
\begin{split}
L_\ell &\mapsto L_\ell' = L_\ell -  J^3_\ell \xi +  \frac{c}{24} \xi^2 \delta_{\ell,0}
\\
J^3_\ell &\mapsto J'^3_\ell = J^3_\ell -  \frac{c}{12} \, \xi \, \delta_{\ell,0}
\end{split}		\label{SpecFlow}
\end{align}
where $\xi$ is the spectral flow parameter. Hence an operator with conformal weight $h$ and R-charge $j^3$ is mapped to an operator with 
\begin{align}
\begin{split}
h' &= h -  j^3 \, \xi + \frac{c}{24}  \xi^2 \, ,
\qquad
j'^3 = j^3 -  \frac{c}{12}  \xi ,
\end{split}		\label{SpecFlowhj}
\end{align}
and NS (anti-)chiral fields flow to Ramond ground states:
\be
h^\NS = \pm j^3 \quad \mapsto \quad h' = \tfrac{1}{24} c 
\quad
\text{for}
\quad
\xi = \pm 1 .
\ee

Our renormalized fields $R^\pm_{[n]}$, which are Ramond ground states of the $n$-wound string, have conformal weight
\be
h^\R_n = \tfrac14 n < \tfrac14 N ,
\ee
the bound $n < N$ being due to our calculation at order $N^{-1}$; the field with $n = N$ scales as $N^{-2}$ for large $N$, and requires a genus-one covering map.
In the free theory, in the presence of a $n$-twist, it is possible to consider the $Z_n$ orbifold $\cal N = 4$ SCFT with $c = 6n$, whose conserved currents  are defined by adding the $n$ copies entering the twist.
For example, taking the cycle to be $(1, \cdots, n)$, the $n$-twisted CFT has stress tensor and Virasoro modes%
	\footnote{%
	In the presence of $\s_{(n)}$ we can also define the usual fractional modes $L_{\frac{k}{n}}$, $J^3_{\frac{k}{n}}$, etc., as well as a `fractional spectral flow' which is an automorphism of the fractional algebra \cite{deBeer:2019ioe}.}
\be	\label{ntwistedT}
T(z) = \sum_{I=1}^n T_I(z) \, ,
\qquad
L_k \; \s_{(1,\cdots,n)}(0) 
	=  \oint \frac{dz}{2\pi i} z^{1+k} \sum_{I =1}^n T_I(z) \s_{(1,\cdots,n)}(0) .
\ee
The  modes $L_k$, with $k \in \mathbb Z$, are well defined: the twist shuffles the terms in the summation over $I$, but the summation itself is preserved. Spectral flow of this $n$-twisted algebra by  $\xi = 1$, when applied to the NS chiral field $O^{(0,0)}_{(1,\cdots,n)}$ with $h^\NS_n = \frac{n-1}{2} = j^3_n$, gives the Ramond field $R^-_{(1,\cdots,n)}$, with $h' = \frac14 n$ and $j'^3 = - \frac12$. Starting with the anti-chiral NS and flowing by $\xi = -1$, we get $R^+_{(1,\cdots,n)}$, etc.

As shown by Eq.(\ref{JOeq0}), the dimension of the field $O^{(0,0)}_{(1,\cdots,n)}$ is protected in the deformed theory, although $R^\pm_{(1,\cdots,n)}$ is renormalized.
How is this to be reconciled with the spectral flow between them?
The answer is that the symmetry algebra of the $n$-twisted CFT, i.e.~the $\cal N = (4,4)$ SCFT with central charge $c = 6n < 6N$, is not preserved after the deformation by the  interaction $\Oint$. 
The basic reason for this is that the twist can join two strings into a longer string. 

Let us discuss this in more detail.
The deformed SCFT has deformed charges $T^{(\la)}(z)$, $J^{(\la) a}(z)$, $G^{(\la)\a A}(z)$, which must close under an operator algebra. If this algebra has an automorphism, then we can define a spectral flow between the deformed states. 
The deformed charges are difficult to describe explicitly. 
In particular, obtaining the stress-tensor $T^{(\la)}(z)$ is subtle, as explained in \cite{Guo:2019pzk}, since one cannot na\"ively make a variation of the action (\ref{defpercft}). Instead, one can use the prescription of Sen\cite{SEN1990551} to obtain the Virasoro modes by the following action on a field $\Phi$, 
\be	\label{laLn}
L_k^{(\la)} \Phi(0) 
	= (L_k + \la \delta L_k) \Phi_0
	=  L_k \Phi(0) + \la \underset{|z|=\e}{\oint} \! d\bar z \ z^{k+1} \, \Oint(z,\bar z) \Phi(0) .
\ee
See also \cite{Guo:2019pzk}, and \cite{Campbell:1990dz} for a discussion of this formula.
As shown in \cite{SEN1990551}, the modes $L^{(\la)}_k$ satisfy the Virasoro algebra with the same central charge as the unperturbed algebra of the $L_k$, which are derived from the unperturbed tensor $T(z)$. Such preservation of the Virasoro algebra after the deformation (\ref{laLn}) requires computing $[L_k , \delta L_l ]$, which in turn includes computing the integral
\be	\label{intToint}
\underset{\text{$z$ around $w$}}{\la \ \oint \! dz} z^{l+1} \, T(z) \Oint(w,\bar w) .
\ee
Now, $\Oint$ is made out of the sum over the conjugacy class of $Z_2$ cycles in $S_N$, hence it involves all the $N$ copies of the orbifold. Therefore the integral above is only defined if $T(z)$ also includes all the $N$ copies of the fields. 
For example, if we take the $n$-twisted algebra made by the modes (\ref{ntwistedT}), when going around the twist $\s_{(n,n+1)}$ the integral would not be defined.
In other words, the Virasoro algebra with $c = 6n$ is inconsistent with the  deformation (\ref{defpercft}) for $n < N$.

	%
	%

Let us also stress another important detail of our computation of the corrected dimension: since we perform the integral $J_R$ by changing coordinates to the covering surface, we automatically include all the conjugacy classes of the permutations inside the four-point function, which are taken into account by the very nature of the covering map, as we have extensively discussed in Sect.\ref{SectOPElimits}. Thus we are truly computing the renormalized fields in the theory deformed by $\Oint$, rather than by a non-$S_N$-invariant operator such as, say, $\Oincov_{(1,2)}$.

Thus spectral flow between $R^\pm_{[n]}$ and $O^{(0,0)}_{[n]}$ is not preserved after the deformation by $\Oint$, unless $n = N$. 
Spectral flow of the full orbifold theory, with $c = 6N$ is, however, (expected to be) preserved.
But, in the full orbifold theory, $R^\pm_{[n]}$ is not a ``true'' Ramond ground state --- it is a mix of a $n$-twisted string in a Ramond ground state, with $N-n$ untwisted strings in the NS ground state. Thus one would not expect $R^\pm_{[n]}$ to be BPS protected. Explicitly, taking $\xi = 1$ and with $c = 6N$, the field $R^-_{[n]}$ with $h^\R_n = \frac14 n$ and $j^3 = - \frac12$ flows to a state with
\be
h' = \frac{ n + N - 2}{4} ,
\qquad
j'^3 = \frac{N-1}{2}
\ee
which is only chiral for $n=N$. 
Conversely, the NS chiral $O^{(0,0)}_{[n]}$ flows to a state with
\be
h' = \tfrac14 N , 
\qquad
j'^3 = - \frac{N - n + 1}{2} 
\ee
which is a ``true'' Ramond ground state made by composing twisted and untwisted R-charged and R-neutral Ramond fields; when $n = N$, this field is simply $R^-_{[N]}$.

Note that this latter state with twist $N$ is indeed protected, as far as our computation is concerned, since, as shown in Eq.(\ref{2nN}), at leading order in $1/N$, the four-point function involving the $R^\pm_{[N]}$ vanishes, and there is no correction to their dimensions.
For the single-cycle fields considered here, this protection is rather trivial, being due simply to the large-$N$ approximation. But we have shown in \cite{Lima:2020nnx} that the protection is again observed in composite fields $R^\a_{[m_1]}R^\b_{[m_2]}(z,\bar z)$ with dimension $h^\R = \frac14 N$, for which $m_1 + m_2 = N$. In this case, there \emph{is} a genus-zero contribution to the four-point function, and protection comes from a non-trivial DF integral being zero. Because of spectral flow, we expect that such results generalize for any composite Ramond field whenever the sum of the composing twists add to $N$.


\section{Discussion}	\label{SectConclusion}

The investigation of the twisted Ramond  sector of  marginally-deformed D1-D5 SCFT$_2$ presented in this paper is based on the  explicit construction of the large-$N$ limit of the four-point function (\ref{4pointR1GuRamon}) of two R-charged Ramond fields and two scalar modulus  operators $\Oint(z, \bar z) = \e_{A  B} G^{-  A}_{-\frac{1}{2}} \tilde G^{ \dot -  B}_{-\frac{1}{2}} O^{(0,0)}_{[2]}(z, \bar z)$.
We have found that $R^\a_{[n]}$ undergoes renormalization for all twists $2 < n <N$, and is protected for the maximal and minimal twist values, $n=N$ and $n=2$. The fields $\s_{[n]}$ also renormalize for $n < N$.
In fact, these four-point functions  provide dynamical information about  \emph{both} theories: the ``free-orbifold point'' SCFT$_2$,  and its marginal deformation, at second order in $\la$.
In what follows, we will briefly address a some open problems whose solutions can eventually be reached by adapting  the methods developed in the present  paper.

\bigskip

\noindent
{\bfseries  More on the properties of non-BPS fields.}
The four-point functions that we have calculated can be used not only for accessing the deformed SCFT$_2$,
but also to give a more complete description of the free orbifold itself.
For example, the OPE data we have extracted from short-distance limits reveal important features of the Ramond sector of the free SCFT$_2$, such as  the conformal weights, R-charges and a few structure constants of the non-BPS twisted Ramond operators  $Y_{n\pm1}^{\pm}$  given by Eq.(\ref{Ydesc}). Their four-point functions with the deformation operator,
$$
\big\langle Y^-_{[m]}(\infty) \Oint(1) \Oint(u,\bar u) Y^+_{[m]}(0) \big\rangle,
$$
can be  explicitly constructed by  the same covering map and the same methods  used here. Computation of this function would provide new relevant CFT data: apart from the corrections to the canonical conformal dimensions (\ref{wheighY}), it also contains, in the corresponding OPE limits, all the super-conformal properties of the next members of the family of the non-BPS twisted Ramond fields.

Relevant information about fuzzball microstates can be extracted from four-point functions similar to (\ref{4ptfuncintro}), but with the deformation operators replaced by NS chiral fields $O_{[2]}^{(p,p)}$, with dimensions $\Delta^{(0,0)}_2=1$ and $\Delta^{(1,1)}_2=2$, viz.
\be
\big\langle R^-_{[n]}(\infty) O^{(p,p)}_{[2]}(1) O^{(p,p)}_{[2]} (u,\bar u) R^+_{[m]}(0) \big\rangle.
\ee
Computing these functions by the methods of Sect.\ref{SectFourPointFuncts} is actually easier than computing (\ref{4ptfuncintro}). 
Their short-distance limits  contain the CFT data --- the conformal dimensions and structure constants --- about the non-BPS fields $X^{(p,p)\pm}_{n\pm1}$ appearing in the OPEs $O^{(p,p)}_{[2] }R_n^{\pm}$. 

\bigskip

\noindent
{\bfseries R-neutral Ramond ground states.}
Here we have focused on the R-charged SU(2)$_{L,R}$ doublets $R^\a_{[n]}$. 
The remaining Ramond fields, that are neutral under R-symmetry and form a doublet of SU(2)$_2$, have only been mentioned in passing. 
The renormalization of such fields with a single cycle, $R^{\dot A}_{[n]}$, can be studied with the same methods of the present paper.
One must compute their four-point function with $\Oint$ by lifting to covering space with the corresponding R-neutral spin fields $S^{\dot A}$, etc.
In practice, the actual computation of the four-point function is more complicated then the one presented in \S\ref{SectStressTensrMth}, because some simplifying cancelations only occur for the R-charged fields.
Once the function is found, however, all the methodology developed here for exploration of operator algebras via short-distance limits, as well as the renormalization scheme, can be applied.
We have presented these results elsewhere \cite{Lima:2020urq}.

\bigskip

\noindent
{\bfseries Lifted vs.~protected states.}
%
Supergravity solutions of fuzzballs correspond to states in the CFT where all component strings are in a Ramond ground state, i.e.~
$\prod_k (R^{\a_k}_{[n_k]})^k$, with $\sum_k k n_k = N$.
If there is only one $n$-wound string, then this state becomes 
$(R^{\b}_{[1]})^{N-n} R^\a_{[n]}$, where the untwisted Ramond field  $R^\b_{[1]}$ is a (symmetrized) spin field. 
The fields that we have found to be renormalized are made by putting the $n$-wound string is in a Ramond ground state, while all the other untwisted strings are in the NS vacuum, i.e.~explicitly $R^\a_{[n]} \cong ({\mathds 1})^{N-n} R^\a_{[n]}$.
The fact that one can define a spectral flow of the $c = 6n$ super-conformal algebra in the $n$-wound string at the free orbifold point may suggest that both $({\mathds 1})^{N-n} R^\a_{[n]}$ and $(R^{\b}_{[1]})^{N-n} R^\a_{[n]}$ are protected. As we have shown here, this is not true.

The single-cycle Ramond field with maximal twist $n=N$, which \emph{is} a ``pure'' Ramond ground state in the full orbifold theory, is, indeed, protected, as far as our analysis goes: the four-point function for this field scales as $1/N^2$ and must be computed with a genus-one covering surface.
The interesting feature that our calculation highlights is that this protection depends crucially on the combinatorics involved in the combinations of the permutation cycles of the twisted fields $R^\a_{[n]}$ and $\Oint$. The protection is due not to the fact that the four-point function or its integral vanish 
--- 
in contrast to what happens in the case of minimal twist $n=2$, where the behavior of the four-point functions changes drastically and $J_R(2)$ vanishes, 
neither the function $z(t)$, the map $u(x)$, the functions $G_R(x)$, nor the integrals $J_R(n)$, none of them is able to distinguish between $2 < n < N$ or $n = N$.
What separates the maximal twist is the analysis of Eq.(\ref{compto1}), which dictates the $N$-dependence of $G_R(u,\bar u)$, and implies that any function obtained from the genus-zero covering map must correspond to a permutation of $S_N$ such that $n < N$.
These features become starker when we consider the composite field 
$
(R^{\a_1}_{[n_1]}R^{\a_2}_{[n_2]}) 
\cong
({\mathds 1})^{N-n_1-n_2} R^{\a_1}_{[n_1]} R^{\a_2}_{[n_2]}
$, 
which was done in \cite{Lima:2020nnx}.
When $n_1 + n_2 < N$, the solutions to the equation equivalent to Eq.(\ref{compto1}) allow for factorizations of
$$
\big\langle 
(R^{\a_1}_{[n_1]}R^{\a_2}_{[n_2]})^\dagger (\infty,\bar \infty) 
\Oint(1,\bar 1) \Oint(u,\bar u) 
(R^{\a_1}_{[n_1]}R^{\a_2}_{[n_2]})(0,\bar 0) \big\rangle
$$
into four-point functions involving only one of the single-cycle fields; then the composite field renormalizes as a corollary of our present results. 
When $n_1 + n_2 = N$, and the field becomes a ``pure'' Ramond ground state with $h = \frac1{24} c_{orb}$, there is no factorization. Now, the covering surface for this completely connected function has genus zero, and we \emph{can} calculate the four-point function and its integral explicitly, in contrast to what happened here for the field $R^\a_{[N]}$. The four-point function we find in \cite{Lima:2020nnx} is non-trivial, and reveals conformal data and OPE fusion rules. Its integral, corresponding to $J_R$, is also non-trivial but \emph{it does vanish} after we apply our regularization procedure and the DF construction: thus we see explicitly that the family of pure Ramond fields $(R^{\a_1}_{[n_1]}R^{\a_2}_{[n_2]})$ with $n_1 + n_2 = N$ is again protected.

In closing, one cannot avoid the question of what are (if any) the bulk holographic images of the renormalized $R^\a_{[n]}$ fields, with their \emph{continuous}, $\la$-dependent conformal dimensions. The answer remains to be discovered, and there are indications that tools necessary for this end include the description  of the symmetry algebra of the deformed SCFT$_2$ and its unitary representations.

\acknowledgments
The work of M.S.~is partially supported by the Bulgarian NSF grant KP-06-H28/5 and that of M.S.~and G.S.~by the Bulgarian NSF grant KP-06-H38/11.
 M.S. is grateful for the kind hospitality of the Federal University of Esp\'irito Santo, Vit\'oria, Brazil, where part of his work was done.

\appendix

\section{Conventions for the $\cal N = (4,4)$ SCFT}
\label{AppConventions}

In the $\cal N = (4,4)$ superalgebra,
the  R-currents $J^a(z)$, $\tilde J^a(\bar z)$, and the supercurrents $G^{\a A} (z)$, $\tilde G^{\dot \a \dot A}(\bar z)$ have indices in SU(2) groups as follows:
$a = 1,2,3$ and $\dot a = \dot 1,\dot2,\dot3$ transform as a triplets of SU(2)$_L$ and SU(2)$_R$, respectively;
$\a = + , -$ and $\dot \a = \dot +, \dot -$ transform as a doublets of SU(2)$_L$ and SU(2)$_R$, respectively; indices $A=1,2$ and $\dot A=\dot1,\dot2$ transform as doublets of SU(2)$_1$ and SU(2)$_2$, respectively.

The SCFT can be realized in terms of four real bosons $X_i(z,\bar z)$, four real holomorphic fermions $\psi_i(z)$ and four real anti-holomorphic fermions $\tilde \psi_i(\bar z)$, with $i =1,\cdots, 4$.
They are related to the complex fields $X^{\dot AA}(z,\bar z)$, $\psi^{\a \dot A}(z)$ and $\tilde \psi^{\dot \a\dot A}(\bar z)$ by
\begin{align}
& X_{\dot A  A} = \tfrac{1}{\surd2} X_i [\s^i]_{\dot A A} = \frac{1}{\surd2} 
				\begin{bmatrix}
				X_3 + i X_4 & X_1 - i X_2
				\\
				X_1 + i X_2 & -X_3 + i X_4
				\end{bmatrix} ,
	\label{XdotAAfromi}
\\
& \psi^{\a \dot 1} =
\begin{bmatrix}
\psi^{+ \dot 1} \\ \psi^{- \dot 1} 
\end{bmatrix}
=
\frac{1}{\surd2}
\begin{bmatrix}
\psi_1+ i \psi_2 \\ \psi_3 + i \psi_4
\end{bmatrix} ,
\quad
\psi^{\a\dot 2} =
\begin{bmatrix}
\psi^{+ \dot 2} \\ \psi^{- \dot 2} 
\end{bmatrix}
=
\frac{1}{\surd2}
\begin{bmatrix}
\psi_3 - i \psi_4 \\ - \psi_1 + i \psi_2
\end{bmatrix}.	 \label{DoublFermiDef}
\end{align}
There are analogous constructions for the right-moving sector.
The Levi-Civita symbol always has the structure $\e^{12} = +1$. Pauli matrices are defined such that $\s^3 = \rm{Diag}(1,-1)$.
The ``Pauli vector''  $\s^i = (\s^1, \s^2, \s^3 , \s^4)$
and its conjugate
 $\bar \s^i$
  have components (we work in Euclidean space)
$\s^a = - \bar \s^a$ $\s^4 = i \mathds 1_{2 \times 2} = \bar \s^4$.

The reality condition of $X_i$ and $\psi_i$ implies that
\be	\label{RealiCondXpsi}
X^{\dot A A} \equiv - \e^{\dot A \dot B} \e^{A B} X_{\dot B B} ,
\quad
( X_{\dot A A} )^\dagger = X^{\dot A A} \ , 
\quad
( \psi^{\a \dot A} )^\dagger = \psi_{\a \dot A} .
\ee
Two-point functions are
\begin{align}
\langle \pa X^{\dot A A}(z) \pa X^{\dot B B} (z') \rangle 
	&=  \frac{2 \e^{\dot A \dot B} \e^{A B}}{(z - z')^2} 	\label{2ptFuncpaXAA}
\\
\langle \psi^{\a \dot A} (z) \psi^{\b \dot B} (z') \rangle 
	&= - \frac{\e^{\a\b} \e^{\dot A \dot B}}{z - z'} 	\label{2ptFuncpsipsi}
\\
\langle \pa \phi_r(z) \pa \phi_s(z') \rangle &= - \frac{\delta_{rs}}{(z - z')^2} 
 \label{2ptFuncXiXi}	
\end{align}
where the last equation is for the bosonized fermions (\ref{Bosnpsi2real}).
The non-vanishing bosonic two-point functions are between a current $\pa X^{\dot A A}$ and its complex conjugate; explicitly,
\be
\langle \pa X^{\dot 1 1}(z) ( \pa X^{\dot 1 1} )^\dagger (z') \rangle = \frac{2 }{(z - z')^2} ,
\quad
\langle \pa X^{\dot 1 2}(z) ( \pa X^{\dot 1 2} )^\dagger (z') \rangle = - \frac{2 }{(z - z')^2} ,
\label{twopntboconj}
\ee
as can be checked from (\ref{2ptFuncpaXAA}) using the reality conditions (\ref{RealiCondXpsi}).

\section{Asymptotics to OPEs}	\label{AppAsympOPE}

In calculating OPEs, we need to know the inverse of (\ref{ux}) near the base-sphere points $u_* = 0, \infty, 1$.
When $u_* = 0$, the roots of Eq.(\ref{RamifEqHu}) are obvious: $x = 0$ (with multiplicity $n-1$) and $x = - n$ (with multiplicity $n+1$). 
Going back to (\ref{ux}), we find the form of $u(x)$ in these two limits,
\begin{align*}
u(x) \approx \frac{n^{n+1}}{(1-n)^{n-1}} x^{n-1} ,
\qquad
u(x) \approx \frac{n^{n-1}}{(-n-1)^{n+1} }  (x + n)^{n+1}
\end{align*}
so inverting we get the two functions
\be
(u \to 0) \quad
	\begin{sqcases}
x \to 0, \quad &x^0_{\frak1}(u) \approx \left( \frac{(1-n)^{n-1}}{n^{n+1}} u\right)^{\frac{1}{n-1}}
\\
x \to -n ,\quad &x^0_{\frak2}(u) \approx - n + \left( \frac{(-n-1)^{n+1} }{n^{n-1}} u \right)^{\frac{1}{n+1}}
\end{sqcases}	\label{xto0minnux}
\ee

Taking $u_* = \infty$, Eq.(\ref{RamifEqHu}) reduces to 
$(x-1)^{n+1}(x+n-1)^{n-1} = 0$,	
with roots $x = 1$ and $x = 1-n$. 
The function $u(x)$ behaves in these limits as
\begin{align*}
u(x) \approx \frac{(1+ n)^{n+1}}{n^{n-1}} \,  \frac{1}{(x-1)^{n+1}} ,
 \qquad 
 u(x) \approx \frac{(n-1)^{n-1}}{n^{n+1} }  \frac{1}{(x -1 + n)^{n-1}}
 \end{align*}
 so we have the inverse functions
\be
(u \to \infty) \quad
	\begin{sqcases}
 x \to 1-n, \quad &x^\infty_{\frak1}(u) \approx 1 - n + \left( \frac{(n-1)^{n-1}}{n^{n+1} } \frac{1}{u} \right)^{\frac{1}{n-1}}
\\
 x \to 1, \quad &x^\infty_{\frak2}(u) \approx 1 + \left( \frac{(1+ n)^{n+1}}{n^{n-1}} \frac{1}{u} \right)^{\frac{1}{n+1}}
\end{sqcases}
	\label{xto0minnuxInf}
\ee

When $u_* =1$, one cannot find the $2n$ solutions of Eq.(\ref{RamifEqHu}), but fortunately we are only interested in those solutions which also correspond to the limit $t_1 \to x$. In this case, instead of a polynomial equation of degree $2n$, we must solve Eq.(\ref{ArtFrolChoice}) which becomes
\be
\frac{2x + n - 1}{(n+x)x}  = 0 	\label{inds3xchans}
\ee
with only two solutions: $x = \infty$ and
$x =  \frac{1}{2}(1-n)$.	
The behavior of $u(x)$ near $x = \infty$ can be found with the conformal transformation $x = 1 / \vare$; evaluating $u(1/\vare)$ around small $\vare$,  
\begin{align*}
u(1/\vare) = 1 + 4n \vare + 2n(1+3n) \vare^2 + \rm{O}(\vare^3)
\end{align*}
while expanding $u(x)$ the second limit, when $x \to \frac{1-n}{2}$, we get
\be
\begin{split}
u(x) &= 1 - \tfrac{64n}{3(n^2-1)^2} \left( x - \tfrac{1-n}{2} \right)^3 
		- \tfrac{512 n (1+n^2)}{5(n^2 -1)^4} \left( x - \tfrac{1-n}{2} \right)^5
\\
		&\qquad + \tfrac{2048 n^2}{9(n^2 -1)^4} \left( x - \tfrac{1-n}{2} \right)^6
+ {\rm O}\left( x - \tfrac{1-n}{2} \right)^7 
\end{split}
	\label{serieuxnear1n2}
\ee
Inverting the two series above, we find
\begin{align}
(u \to 1) \quad
	\begin{sqcases}
x \to \infty , &\quad x^1_{\frak1}(u) \approx -{4n\over 1-u}+ \frac{3n+1}{2}+ \cdots
\\
\\
 x \to \tfrac{1-n}{2} , &\quad x^1_{\frak2}(u) \approx \frac{1-n}{2} 
\\
&\quad\quad\quad\quad	
				+ \tfrac{3^{1/3} (n^2 -1)^{2/3}}{4n^{1/3}} (1 - u)^{\frac{1}{3}} 
\\
&\quad\quad\quad\quad	
				- \tfrac{3(n^2 +1)}{40n} (1-u) 
\\				
&\quad\quad\quad\quad	
				+\tfrac{ (n^2 -1)^{2/3}}{8 \cdot 3^{2/3} \cdot n^{1/3}} (1 -u)^{\frac{4}{3}} 
				+ \cdots
	\end{sqcases} \label{xaxp4ns}
\end{align}
Note that the multiplicity of the solution $x = \frac{1-n}{2}$ is 3, and that of $x = \infty$ is 1.

\section{OPEs with bare twists and structure constants} \label{OPEsss} \label{SectStrucConst} 	\label{SectOPEsss}

In this appendix, we examine the OPE limits of the functions $G_\s(x)$ and $g(x)$. We derive several structure constants, some of which are known in the literature, thus checking our expressions for $g(x)$ and $G_\s(x)$. 

\bigskip

We start with the limit $u \to 1$ for $G(u,\bar u)$. The identity channel is the same as for the Ramond fields discussed in the text, while the second channel gives
\begin{align}
\begin{split}
G_\s(x^1_{\frak2}(u)) &= 
		\frac{d_4}{(1-u)^{4/3}}
		+ \frac{d_2}{(1-u)^{2/3}} + \frac{d_1}{(1-u)^{1/3}} + \text{non-sing.}
\end{split}
\end{align}	
where, after taking (\ref{CRCs}) into account,
\begin{align}
\begin{split}
\log |d_4|^2 &= \left( n + \frac{1}{n} + \frac{2}{3} \right) \log (n+1) 
			- \left( n + \frac{1}{n} - \frac{2}{3} \right) \log(n-1) 
			- \frac{4}{3} \log n
\\
		&\quad
			- 4 \log 2 - \frac{8}{3} \log 3 .
\label{d4log}			
\end{split}\end{align}
The powers of $u$ reveal the conformal family of $\s_3$ with no  operator of dimension one among the descendants.
Inserting the OPE (\ref{OPEOinOins3}) back into the four-point function, we find that
\begin{align}
|d_4|^2 = \big\langle \Oincov_2(\infty) \s_3(1) \Oincov_2(0) \big\rangle
		\, \big\langle \s_n(\infty) \s_3(1) \s_n(0) \big\rangle	.\label{3strcConsOO}
\end{align}
The structure constant $C^{\Oincov \s \Oincov}_{232} =  \langle \Oincov_2(\infty) \s_3(1) \Oincov_2(0) \rangle$ carries no $n$-dependence,
and we can write $C^\s_{n3n} = \langle \s_n(\infty) \s_3(1) \s_n(0) \rangle$ as
\be
\log C^\s_{n3n}  = \left( n + \frac{1}{n} + \frac{2}{3} \right) \log (n+1) 
			- \left( n + \frac{1}{n} - \frac{2}{3} \right) \log(n-1) 
			- \frac{4}{3} \log n
			+ \kappa ,
		\label{Csn3n}	
\ee
which agrees with Eq.(6.25) of Ref.\cite{Lunin:2000yv}, apart from an overall factor of $\frac{1}{6}$.
The $n$-independent number
$\kappa = \tfrac{4}{3} \log3 + \tfrac{1}{3}\log2$ 
can be obtained by taking $n=2$ in Eq.(\ref{Csn3n}), and comparing with Eq.(\ref{Csn2n1m}), below. Hence
\be
\log C^{\Oincov \s \Oincov}_{232} = 4 \log3 + \tfrac{13}{3} \log2 .
\label{Csint232}
\ee

An interesting check of the results above comes from the function (\ref{functwists}), whose limit $u \to 1$ now gives the OPE $\s_2(u) \s_2(1)$. 
Counting powers of $u$ in
\begin{align}
g(x^1_{\frak1}(u)) &=  \frac{ c_\s (- 4 n)^{3/4}}{(1-u)^{3/4}} + \rm{O}(1-u)^{1/4}	\label{gx11}
\\
g(x^1_{\frak2}(u)) &= \frac{b}{(1-u)^{1\over12}} + \rm{O}(1-u)^{7/12}	\label{gx12}
\end{align}
 Eq.(\ref{gx11}) gives again the identity,  thus determining 
$c_\s = (-4n)^{-3/4}$.
In the channel (\ref{gx12}) we find $\s_3$ with its dimension 
$h^\s_3 = \frac{2}{3}$, 
completing the well-known fusion rule
$[\s_2] \times [\s_2] = [\s_1] + [\s_3]$. Note the absence of next-to leading singularities in Eqs.(\ref{gx11}) and (\ref{gx12}); there are no descendants in these OPEs.

The constant $b$ gives information about $\langle \s_p \s_q \s_r \rangle =  C^\s_{pqr}$:
\begin{align}
\log C^\s_{n3n} + \log C^\s_{232} 
	&= \log |b|^2 
\nonumber
\\
	&= \left( n + \frac{1}{n} + \frac{2}{3} \right) \log(n+1) 
		- \left( n + \frac{1}{n} - \frac{2}{3} \right) \log(n-1)
\nonumber
\\
&\qquad		- \frac{4}{3} \log n
			- 4 \log 2
			- \frac{1}{6} \log 3	
\end{align}
Comparison with Eqs.(\ref{3strcConsOO}) and (\ref{d4log})  reveals the same $n$-dependence for both three-point functions --- an important cross-check between the two  $g(x)$ and $G_\s(x)$ (which were obtained independently).


The more general fusion rule
\be
[\sigma_2] \times [\sigma_n] = [\sigma_{n-1}] + [\sigma_{n+1}]	\label{fusionsimnn}
\ee
can be derived from 
\begin{align*}
g(x^0_{\frak1}(u)) = u^{- \frac{5n^2 - 5n + 2}{8n(n-1)}} \big( c_1^- + c_2^- u^{\frac{1}{n-1}} + \cdots \big) ,
\quad
g(x^0_{\frak2}(u)) = u^{- \frac{5n^2 + 5n - 2}{8n(n+1)}} \big( c_1^+ + c_2^+ u^{\frac{1}{n+1}} + \cdots \big)
\end{align*}
where the coefficients $c_1^\pm$ are readily computable.
One can check from the powers of $u$ that channels $x^0_{\frak1}(u)$ and $x^0_{\frak2}(u)$ give operators of dimensions $h^\s_{n-1}$ and $h^\s_{n+1}$, that is $\s_{n-1}$ and $\s_{n+1}$, respectively. 
The coefficients $c_1^\pm$ give us information about 
$\big\langle \s_n(\infty) \s_2(1) \s_{n-1}(0) \big\rangle = |c_1^-|$ 
and
$ \big\langle \s_n(\infty) \s_2(1) \s_{n+1}(0) \big\rangle = |c_1^+|$. 
Explicitly
\begin{align}
\log C^\s_{n,2,n-1} &= - \frac{2n^2 - n + 2}{4n} \log(n-1) 
			+ \frac{2n^2 - 3n + 3}{4(n-1)} \log n
			- \frac{5}{4} \log 2		\label{Csn2n1m}
\\
\log C^\s_{n,2,n+1} &= + \frac{2n^2 +n + 2}{4n} \log(n+1) 
			- \frac{2n^2 + 3n + 3}{4n(n+1)}  \log n
			- \frac{5}{4} \log 2		\label{Csn2n1p}
\end{align}
This agrees with the result of Ref.\cite{Lunin:2000yv}, again, apart from an overall factor of $\frac{1}{6}$.
Inserting $n=3$ in Eq.(\ref{Csn2n1m}), we find that 
$\log C^\s_{3,2,2} = - \frac{5}{3} \log 2$, which was used to derived Eq.(\ref{Csint232}).

\bibliographystyle{utphys}

\bibliography{D1D5LongPaperReferencesV40} 

\providecommand{\href}[2]{#2}\begingroup\raggedright\begin{thebibliography}{10}

\bibitem{Maldacena:1997re}
J.~M. Maldacena, ``{The Large N limit of superconformal field theories and
  supergravity},'' \href{http://dx.doi.org/10.1023/A:1026654312961}{{\em Int.
  J. Theor. Phys.} {\bf 38} (1999)  1113--1133},
  \href{http://arxiv.org/abs/hep-th/9711200}{{\tt arXiv:hep-th/9711200}}.

\bibitem{Maldacena:1998bw}
J.~M. Maldacena and A.~Strominger, ``{AdS(3) black holes and a stringy
  exclusion principle},''
  \href{http://dx.doi.org/10.1088/1126-6708/1998/12/005}{{\em JHEP} {\bf 12}
  (1998)  005}, \href{http://arxiv.org/abs/hep-th/9804085}{{\tt
  arXiv:hep-th/9804085}}.

\bibitem{Seiberg:1999xz}
N.~Seiberg and E.~Witten, ``{The D1 / D5 system and singular CFT},''
  \href{http://dx.doi.org/10.1088/1126-6708/1999/04/017}{{\em JHEP} {\bf 04}
  (1999)  017}, \href{http://arxiv.org/abs/hep-th/9903224}{{\tt
  arXiv:hep-th/9903224}}.

\bibitem{David:2002wn}
J.~R. David, G.~Mandal, and S.~R. Wadia, ``{Microscopic formulation of black
  holes in string theory},''
  \href{http://dx.doi.org/10.1016/S0370-1573(02)00271-5}{{\em Phys. Rept.} {\bf
  369} (2002)  549--686}, \href{http://arxiv.org/abs/hep-th/0203048}{{\tt
  arXiv:hep-th/0203048}}.

\bibitem{Mathur:2005zp}
S.~D. Mathur, ``{The Fuzzball proposal for black holes: An Elementary
  review},'' \href{http://dx.doi.org/10.1002/prop.200410203}{{\em Fortsch.
  Phys.} {\bf 53} (2005)  793--827},
  \href{http://arxiv.org/abs/hep-th/0502050}{{\tt arXiv:hep-th/0502050}}.

\bibitem{Skenderis:2008qn}
K.~Skenderis and M.~Taylor, ``{The fuzzball proposal for black holes},''
  \href{http://dx.doi.org/10.1016/j.physrep.2008.08.001}{{\em Phys. Rept.} {\bf
  467} (2008)  117--171}, \href{http://arxiv.org/abs/0804.0552}{{\tt
  arXiv:0804.0552 [hep-th]}}.

\bibitem{Strominger:1996sh}
A.~Strominger and C.~Vafa, ``{Microscopic origin of the Bekenstein-Hawking
  entropy},'' \href{http://dx.doi.org/10.1016/0370-2693(96)00345-0}{{\em Phys.
  Lett. B} {\bf 379} (1996)  99--104},
  \href{http://arxiv.org/abs/hep-th/9601029}{{\tt arXiv:hep-th/9601029}}.

\bibitem{Lunin:2001jy}
O.~Lunin and S.~D. Mathur, ``{AdS / CFT duality and the black hole information
  paradox},'' \href{http://dx.doi.org/10.1016/S0550-3213(01)00620-4}{{\em Nucl.
  Phys. B} {\bf 623} (2002)  342--394},
  \href{http://arxiv.org/abs/hep-th/0109154}{{\tt arXiv:hep-th/0109154}}.

\bibitem{Kanitscheider:2007wq}
I.~Kanitscheider, K.~Skenderis, and M.~Taylor, ``{Fuzzballs with internal
  excitations},'' \href{http://dx.doi.org/10.1088/1126-6708/2007/06/056}{{\em
  JHEP} {\bf 06} (2007)  056}, \href{http://arxiv.org/abs/0704.0690}{{\tt
  arXiv:0704.0690 [hep-th]}}.

\bibitem{Kanitscheider:2006zf}
I.~Kanitscheider, K.~Skenderis, and M.~Taylor, ``{Holographic anatomy of
  fuzzballs},'' \href{http://dx.doi.org/10.1088/1126-6708/2007/04/023}{{\em
  JHEP} {\bf 04} (2007)  023}, \href{http://arxiv.org/abs/hep-th/0611171}{{\tt
  arXiv:hep-th/0611171}}.

\bibitem{Mathur:2018tib}
S.~D. Mathur and D.~Turton, ``{The fuzzball nature of two-charge black hole
  microstates},'' \href{http://dx.doi.org/10.1016/j.nuclphysb.2019.114684}{{\em
  Nucl. Phys. B} {\bf 945} (2019)  114684},
  \href{http://arxiv.org/abs/1811.09647}{{\tt arXiv:1811.09647 [hep-th]}}.

\bibitem{Lunin:2000yv}
O.~Lunin and S.~D. Mathur, ``{Correlation functions for M**N / S(N)
  orbifolds},'' \href{http://dx.doi.org/10.1007/s002200100431}{{\em Commun.
  Math. Phys.} {\bf 219} (2001)  399--442},
\href{http://arxiv.org/abs/hep-th/0006196}{{\tt arXiv:hep-th/0006196
  [hep-th]}}.

\bibitem{Lunin:2001pw}
O.~Lunin and S.~D. Mathur, ``{Three point functions for $M^N / S_N$ orbifolds
  with ${\mathcal N}=4$ supersymmetry},''
  \href{http://dx.doi.org/10.1007/s002200200638}{{\em Commun. Math. Phys.} {\bf
  227} (2002)  385--419},
\href{http://arxiv.org/abs/hep-th/0103169}{{\tt arXiv:hep-th/0103169
  [hep-th]}}.

\bibitem{Balasubramanian:2005qu}
V.~Balasubramanian, P.~Kraus, and M.~Shigemori, ``{Massless black holes and
  black rings as effective geometries of the D1-D5 system},''
  \href{http://dx.doi.org/10.1088/0264-9381/22/22/010}{{\em Class. Quant.
  Grav.} {\bf 22} (2005)  4803--4838},
  \href{http://arxiv.org/abs/hep-th/0508110}{{\tt arXiv:hep-th/0508110}}.

\bibitem{Avery:2009tu}
S.~G. Avery, B.~D. Chowdhury, and S.~D. Mathur, ``{Emission from the D1D5
  CFT},'' \href{http://dx.doi.org/10.1088/1126-6708/2009/10/065}{{\em JHEP}
  {\bf 10} (2009)  065}, \href{http://arxiv.org/abs/0906.2015}{{\tt
  arXiv:0906.2015 [hep-th]}}.

\bibitem{Pakman:2009ab}
A.~Pakman, L.~Rastelli, and S.~S. Razamat, ``{Extremal Correlators and Hurwitz
  Numbers in Symmetric Product Orbifolds},''
  \href{http://dx.doi.org/10.1103/PhysRevD.80.086009}{{\em Phys. Rev.} {\bf
  D80} (2009)  086009},
\href{http://arxiv.org/abs/0905.3451}{{\tt arXiv:0905.3451 [hep-th]}}.

\bibitem{Pakman:2009mi}
A.~Pakman, L.~Rastelli, and S.~S. Razamat, ``{A Spin Chain for the Symmetric
  Product CFT(2)},'' \href{http://dx.doi.org/10.1007/JHEP05(2010)099}{{\em
  JHEP} {\bf 05} (2010)  099},
\href{http://arxiv.org/abs/0912.0959}{{\tt arXiv:0912.0959 [hep-th]}}.

\bibitem{Burrington:2012yq}
B.~A. Burrington, A.~W. Peet, and I.~G. Zadeh, ``{Operator mixing for string
  states in the D1-D5 CFT near the orbifold point},''
  \href{http://dx.doi.org/10.1103/PhysRevD.87.106001}{{\em Phys. Rev.} {\bf
  D87} (2013) no.~10, 106001},
\href{http://arxiv.org/abs/1211.6699}{{\tt arXiv:1211.6699 [hep-th]}}.

\bibitem{Bena:2013dka}
I.~Bena and N.~P. Warner, ``{Resolving the Structure of Black Holes:
  Philosophizing with a Hammer },'' \href{http://arxiv.org/abs/1311.4538}{{\tt
  arXiv:1311.4538 [hep-th]}}.

\bibitem{Carson:2014ena}
Z.~Carson, S.~Hampton, S.~D. Mathur, and D.~Turton, ``{Effect of the
  deformation operator in the D1D5 CFT},''
  \href{http://dx.doi.org/10.1007/JHEP01(2015)071}{{\em JHEP} {\bf 01} (2015)
  071}, \href{http://arxiv.org/abs/1410.4543}{{\tt arXiv:1410.4543 [hep-th]}}.

\bibitem{Carson:2015ohj}
Z.~Carson, S.~Hampton, and S.~D. Mathur, ``{Second order effect of twist
  deformations in the D1D5 CFT},''
  \href{http://dx.doi.org/10.1007/JHEP04(2016)115}{{\em JHEP} {\bf 04} (2016)
  115}, \href{http://arxiv.org/abs/1511.04046}{{\tt arXiv:1511.04046
  [hep-th]}}.

\bibitem{Fitzpatrick:2016ive}
A.~L. Fitzpatrick, J.~Kaplan, D.~Li, and J.~Wang, ``{On information loss in
  AdS$_{3}$/CFT$_{2}$},'' \href{http://dx.doi.org/10.1007/JHEP05(2016)109}{{\em
  JHEP} {\bf 05} (2016)  109}, \href{http://arxiv.org/abs/1603.08925}{{\tt
  arXiv:1603.08925 [hep-th]}}.

\bibitem{Burrington:2017jhh}
B.~A. Burrington, I.~T. Jardine, and A.~W. Peet, ``{Operator mixing in deformed
  D1D5 CFT and the OPE on the cover},''
  \href{http://dx.doi.org/10.1007/JHEP06(2017)149}{{\em JHEP} {\bf 06} (2017)
  149}, \href{http://arxiv.org/abs/1703.04744}{{\tt arXiv:1703.04744
  [hep-th]}}.

\bibitem{Galliani:2017jlg}
A.~Galliani, S.~Giusto, and R.~Russo, ``{Holographic 4-point correlators with
  heavy states},'' \href{http://dx.doi.org/10.1007/JHEP10(2017)040}{{\em JHEP}
  {\bf 10} (2017)  040}, \href{http://arxiv.org/abs/1705.09250}{{\tt
  arXiv:1705.09250 [hep-th]}}.

\bibitem{Bombini:2017sge}
A.~Bombini, A.~Galliani, S.~Giusto, E.~Moscato, and R.~Russo, ``{Unitary
  4-point correlators from classical geometries},''
  \href{http://dx.doi.org/10.1140/epjc/s10052-017-5492-3}{{\em Eur. Phys. J. C}
  {\bf 78} (2018) no.~1, 8}, \href{http://arxiv.org/abs/1710.06820}{{\tt
  arXiv:1710.06820 [hep-th]}}.

\bibitem{Tormo:2018fnt}
J.~Garcia~i Tormo and M.~Taylor, ``{Correlation functions in the D1-D5 orbifold
  CFT},'' \href{http://dx.doi.org/10.1007/JHEP06(2018)012}{{\em JHEP} {\bf 06}
  (2018)  012}, \href{http://arxiv.org/abs/1804.10205}{{\tt arXiv:1804.10205
  [hep-th]}}.

\bibitem{Eberhardt:2018ouy}
L.~Eberhardt, M.~R. Gaberdiel, and R.~Gopakumar, ``{The Worldsheet Dual of the
  Symmetric Product CFT},''
  \href{http://dx.doi.org/10.1007/JHEP04(2019)103}{{\em JHEP} {\bf 04} (2019)
  103}, \href{http://arxiv.org/abs/1812.01007}{{\tt arXiv:1812.01007
  [hep-th]}}.

\bibitem{Bena:2019azk}
I.~Bena, P.~Heidmann, R.~Monten, and N.~P. Warner, ``{Thermal Decay without
  Information Loss in Horizonless Microstate Geometries},''
  \href{http://dx.doi.org/10.21468/SciPostPhys.7.5.063}{{\em SciPost Phys.}
  {\bf 7} (2019) no.~5, 063}, \href{http://arxiv.org/abs/1905.05194}{{\tt
  arXiv:1905.05194 [hep-th]}}.

\bibitem{Dei:2019osr}
A.~Dei, L.~Eberhardt, and M.~R. Gaberdiel, ``{Three-point functions in
  AdS$_{3}$/CFT$_{2}$ holography},''
  \href{http://dx.doi.org/10.1007/JHEP12(2019)012}{{\em JHEP} {\bf 12} (2019)
  012}, \href{http://arxiv.org/abs/1907.13144}{{\tt arXiv:1907.13144
  [hep-th]}}.

\bibitem{Eberhardt:2019ywk}
L.~Eberhardt, M.~R. Gaberdiel, and R.~Gopakumar, ``{Deriving the
  AdS$_{3}$/CFT$_{2}$ correspondence},''
  \href{http://dx.doi.org/10.1007/JHEP02(2020)136}{{\em JHEP} {\bf 02} (2020)
  136}, \href{http://arxiv.org/abs/1911.00378}{{\tt arXiv:1911.00378
  [hep-th]}}.

\bibitem{Giusto:2018ovt}
S.~Giusto, R.~Russo, and C.~Wen, ``{Holographic correlators in AdS$_{3}$},''
  \href{http://dx.doi.org/10.1007/JHEP03(2019)096}{{\em JHEP} {\bf 03} (2019)
  096}, \href{http://arxiv.org/abs/1812.06479}{{\tt arXiv:1812.06479
  [hep-th]}}.

\bibitem{Martinec:2019wzw}
E.~J. Martinec, S.~Massai, and D.~Turton, ``{Little Strings, Long Strings, and
  Fuzzballs},'' \href{http://dx.doi.org/10.1007/JHEP11(2019)019}{{\em JHEP}
  {\bf 11} (2019)  019}, \href{http://arxiv.org/abs/1906.11473}{{\tt
  arXiv:1906.11473 [hep-th]}}.

\bibitem{Hampton:2019csz}
S.~Hampton and S.~D. Mathur, ``{Thermalization in the D1D5 CFT },''
  \href{http://arxiv.org/abs/1910.01690}{{\tt arXiv:1910.01690 [hep-th]}}.

\bibitem{Warner:2019jll}
N.~P. Warner, ``{Lectures on Microstate Geometries },''
  \href{http://arxiv.org/abs/1912.13108}{{\tt arXiv:1912.13108 [hep-th]}}.

\bibitem{Dei:2019iym}
A.~Dei and L.~Eberhardt, ``{Correlators of the symmetric product orbifold},''
  \href{http://dx.doi.org/10.1007/JHEP01(2020)108}{{\em JHEP} {\bf 01} (2020)
  108}, \href{http://arxiv.org/abs/1911.08485}{{\tt arXiv:1911.08485
  [hep-th]}}.

\bibitem{Galliani:2016cai}
A.~Galliani, S.~Giusto, E.~Moscato, and R.~Russo, ``{Correlators at large c
  without information loss},''
  \href{http://dx.doi.org/10.1007/JHEP09(2016)065}{{\em JHEP} {\bf 09} (2016)
  065}, \href{http://arxiv.org/abs/1606.01119}{{\tt arXiv:1606.01119
  [hep-th]}}.

\bibitem{Bombini:2019vnc}
A.~Bombini and A.~Galliani, ``{AdS$_{3}$ four-point functions from $
  \frac{1}{8} $ -BPS states},''
  \href{http://dx.doi.org/10.1007/JHEP06(2019)044}{{\em JHEP} {\bf 06} (2019)
  044}, \href{http://arxiv.org/abs/1904.02656}{{\tt arXiv:1904.02656
  [hep-th]}}.

\bibitem{Giusto:2020mup}
S.~Giusto, M.~R. Hughes, and R.~Russo, ``{The Regge limit of AdS$_3$
  holographic correlators},'' \href{http://arxiv.org/abs/2007.12118}{{\tt
  arXiv:2007.12118 [hep-th]}}.

\bibitem{Guo:2019pzk}
B.~Guo and S.~D. Mathur, ``{Lifting of states in 2-dimensional $N = 4$
  supersymmetric CFTs},'' \href{http://dx.doi.org/10.1007/JHEP10(2019)155}{{\em
  JHEP} {\bf 10} (2019)  155}, \href{http://arxiv.org/abs/1905.11923}{{\tt
  arXiv:1905.11923 [hep-th]}}.

\bibitem{Keller:2019suk}
C.~A. Keller and I.~G. Zadeh, ``{Lifting 1/4-BPS States on K3 and Mathieu
  Moonshine },'' \href{http://arxiv.org/abs/1905.00035}{{\tt arXiv:1905.00035
  [hep-th]}}.

\bibitem{Keller:2019yrr}
C.~A. Keller and I.~G. Zadeh, ``{Conformal Perturbation Theory for Twisted
  Fields},'' \href{http://dx.doi.org/10.1088/1751-8121/ab6b91}{{\em J. Phys. A}
  {\bf 53} (2020) no.~9, 095401}, \href{http://arxiv.org/abs/1907.08207}{{\tt
  arXiv:1907.08207 [hep-th]}}.

\bibitem{Guo:2019ady}
B.~Guo and S.~D. Mathur, ``{Lifting of level-1 states in the D1D5 CFT},''
  \href{http://dx.doi.org/10.1007/JHEP03(2020)028}{{\em JHEP} {\bf 03} (2020)
  028}, \href{http://arxiv.org/abs/1912.05567}{{\tt arXiv:1912.05567
  [hep-th]}}.

\bibitem{Belin:2019rba}
A.~Belin, A.~Castro, C.~A. Keller, and B.~M{\"u}hlmann, ``{The Holographic
  Landscape of Symmetric Product Orbifolds},''
  \href{http://dx.doi.org/10.1007/JHEP01(2020)111}{{\em JHEP} {\bf 01} (2020)
  111}, \href{http://arxiv.org/abs/1910.05342}{{\tt arXiv:1910.05342
  [hep-th]}}.

\bibitem{Guo:2020gxm}
B.~Guo and S.~D. Mathur, ``{Lifting at higher levels in the D1D5 CFT},''
  \href{http://arxiv.org/abs/2008.01274}{{\tt arXiv:2008.01274 [hep-th]}}.

\bibitem{Avery:2010er}
S.~G. Avery, B.~D. Chowdhury, and S.~D. Mathur, ``{Deforming the D1D5 CFT away
  from the orbifold point},''
  \href{http://dx.doi.org/10.1007/JHEP06(2010)031}{{\em JHEP} {\bf 06} (2010)
  031}, \href{http://arxiv.org/abs/1002.3132}{{\tt arXiv:1002.3132 [hep-th]}}.

\bibitem{Avery:2010hs}
S.~G. Avery, B.~D. Chowdhury, and S.~D. Mathur, ``{Excitations in the deformed
  D1D5 CFT},'' \href{http://dx.doi.org/10.1007/JHEP06(2010)032}{{\em JHEP} {\bf
  06} (2010)  032}, \href{http://arxiv.org/abs/1003.2746}{{\tt arXiv:1003.2746
  [hep-th]}}.

\bibitem{Carson:2016uwf}
Z.~Carson, S.~Hampton, and S.~D. Mathur, ``{Full action of two deformation
  operators in the D1D5 CFT},''
  \href{http://dx.doi.org/10.1007/JHEP11(2017)096}{{\em JHEP} {\bf 11} (2017)
  096}, \href{http://arxiv.org/abs/1612.03886}{{\tt arXiv:1612.03886
  [hep-th]}}.

\bibitem{Lima:2020boh}
A.~Lima, G.~Sotkov, and M.~Stanishkov, ``{Microstate Renormalization in
  Deformed D1-D5 SCFT},''
  \href{http://dx.doi.org/10.1016/j.physletb.2020.135630}{{\em Phys. Lett. B}
  {\bf 808} (2020)  135630}, \href{http://arxiv.org/abs/2005.06702}{{\tt
  arXiv:2005.06702 [hep-th]}}.

\bibitem{Dixon:1985jw}
L.~J. Dixon, J.~A. Harvey, C.~Vafa, and E.~Witten, ``{Strings on Orbifolds},''
  \href{http://dx.doi.org/10.1016/0550-3213(85)90593-0}{{\em Nucl. Phys. B}
  {\bf 261} (1985)  678--686}.

\bibitem{Arutyunov:1997gi}
G.~Arutyunov and S.~Frolov, ``{Four graviton scattering amplitude from S**N
  R**8 supersymmetric orbifold sigma model},''
  \href{http://dx.doi.org/10.1016/S0550-3213(98)00326-5}{{\em Nucl. Phys. B}
  {\bf 524} (1998)  159--206}, \href{http://arxiv.org/abs/hep-th/9712061}{{\tt
  arXiv:hep-th/9712061}}.

\bibitem{Arutyunov:1997gt}
G.~E. Arutyunov and S.~A. Frolov, ``{Virasoro amplitude from the S**N R**24
  orbifold sigma model},'' \href{http://dx.doi.org/10.1007/BF02557107}{{\em
  Theor. Math. Phys.} {\bf 114} (1998)  43--66},
\href{http://arxiv.org/abs/hep-th/9708129}{{\tt arXiv:hep-th/9708129
  [hep-th]}}.

\bibitem{Pakman:2009zz}
A.~Pakman, L.~Rastelli, and S.~S. Razamat, ``{Diagrams for Symmetric Product
  Orbifolds},'' \href{http://dx.doi.org/10.1088/1126-6708/2009/10/034}{{\em
  JHEP} {\bf 10} (2009)  034},
\href{http://arxiv.org/abs/0905.3448}{{\tt arXiv:0905.3448 [hep-th]}}.

\bibitem{Dotsenko:1984nm}
V.~S. Dotsenko and V.~A. Fateev, ``{Conformal Algebra and Multipoint
  Correlation Functions in Two-Dimensional Statistical Models},''
  \href{http://dx.doi.org/10.1016/0550-3213(84)90269-4}{{\em Nucl. Phys.} {\bf
  B240} (1984)  312}.
[,653(1984)].

\bibitem{Dotsenko:1984ad}
V.~S. Dotsenko and V.~A. Fateev, ``{Four Point Correlation Functions and the
  Operator Algebra in the Two-Dimensional Conformal Invariant Theories with the
  Central Charge c < 1},''
\href{http://dx.doi.org/10.1016/S0550-3213(85)80004-3}{{\em Nucl. Phys.} {\bf
  B251} (1985)  691--734}.

\bibitem{dotsenko1988lectures}
V.~S. Dotsenko, ``Lectures on conformal field theory,'' in {\em Conformal Field
  Theory and Solvable Lattice Models}, pp.~123--170, Mathematical Society of
  Japan.
\newblock 1988.

\bibitem{Mussardo:1987eq}
G.~Mussardo, G.~Sotkov, and M.~Stanishkov, ``{Ramond Sector of the
  Supersymmetric Minimal Models},''
  \href{http://dx.doi.org/10.1016/0370-2693(87)90038-4}{{\em Phys. Lett. B}
  {\bf 195} (1987)  397--406}.

\bibitem{Mussardo:1987ab}
G.~Mussardo, G.~Sotkov, and H.~Stanishkov, ``{Fine Structure of the
  Supersymmetric Operator Product Expansion Algebras},''
  \href{http://dx.doi.org/10.1016/0550-3213(88)90686-4}{{\em Nucl. Phys. B}
  {\bf 305} (1988)  69--108}.

\bibitem{Mussardo:1988av}
G.~Mussardo, G.~Sotkov, and M.~Stanishkov, ``{N=2 SUPERCONFORMAL MINIMAL
  MODELS},'' \href{http://dx.doi.org/10.1142/S0217751X89000522}{{\em Int. J.
  Mod. Phys. A} {\bf 4} (1989)  1135}.

\bibitem{Lima:2020nnx}
A.~Lima, G.~Sotkov, and M.~Stanishkov, ``{Correlation functions of composite
  Ramond fields in deformed D1-D5 orbifold SCFT$_2$},''
  \href{http://arxiv.org/abs/2006.16303}{{\tt arXiv:2006.16303 [hep-th]}}.

\bibitem{Burrington:2018upk}
B.~A. Burrington, I.~T. Jardine, and A.~W. Peet, ``{The OPE of bare twist
  operators in bosonic $S_N$ orbifold CFTs at large $N$},''
  \href{http://dx.doi.org/10.1007/JHEP08(2018)202}{{\em JHEP} {\bf 08} (2018)
  202},
\href{http://arxiv.org/abs/1804.01562}{{\tt arXiv:1804.01562 [hep-th]}}.

\bibitem{deBeer:2019ioe}
T.~De~Beer, B.~A. Burrington, I.~T. Jardine, and A.~W. Peet, ``{The large $N$
  limit of OPEs in symmetric orbifold CFTs with $\mathcal{N}=(4,4)$
  supersymmetry},'' \href{http://dx.doi.org/10.1007/s13130-019-11019-2}{{\em
  JHEP} {\bf 08} (2019)  015},
\href{http://arxiv.org/abs/1904.07816}{{\tt arXiv:1904.07816 [hep-th]}}.

\bibitem{Dixon:1986qv}
L.~J. Dixon, D.~Friedan, E.~J. Martinec, and S.~H. Shenker, ``{The Conformal
  Field Theory of Orbifolds},''
\href{http://dx.doi.org/10.1016/0550-3213(87)90676-6}{{\em Nucl. Phys.} {\bf
  B282} (1987)  13--73}.

\bibitem{Dijkgraaf:1996xw}
R.~Dijkgraaf, G.~W. Moore, E.~P. Verlinde, and H.~L. Verlinde, ``{Elliptic
  genera of symmetric products and second quantized strings},''
  \href{http://dx.doi.org/10.1007/s002200050087}{{\em Commun. Math. Phys.} {\bf
  185} (1997)  197--209}, \href{http://arxiv.org/abs/hep-th/9608096}{{\tt
  arXiv:hep-th/9608096}}.

\bibitem{Roumpedakis:2018tdb}
K.~Roumpedakis, ``{Comments on the S$_{N}$ orbifold CFT in the large
  $N$-limit},'' \href{http://dx.doi.org/10.1007/JHEP07(2018)038}{{\em JHEP}
  {\bf 07} (2018)  038}, \href{http://arxiv.org/abs/1804.03207}{{\tt
  arXiv:1804.03207 [hep-th]}}.

\bibitem{lando2013graphs}
S.~K. Lando and A.~K. Zvonkin, {\em Graphs on surfaces and their applications},
  vol.~141.
\newblock Springer Science \& Business Media, 2013.

\bibitem{NIST:DLMF151}
``{\it NIST Digital Library of Mathematical Functions, \S15.1}.''
  Http://dlmf.nist.gov/, release 1.0.24 of 2019-09-15.
\newblock \url{http://dlmf.nist.gov/15.1}. F.~W.~J. Olver, A.~B. {Olde
  Daalhuis}, D.~W. Lozier, B.~I. Schneider, R.~F. Boisvert, C.~W. Clark, B.~R.
  Miller, B.~V. Saunders, H.~S. Cohl, and M.~A. McClain, eds.

\bibitem{bateman1953higher}
H.~Bateman, {\em {Higher Transcendental Functions}}, vol.~{I-III}.
\newblock McGraw-Hill Book Company, 1953.

\bibitem{NIST:DLMF}
``{\it NIST Digital Library of Mathematical Functions}.''
  Http://dlmf.nist.gov/, release 1.0.24 of 2019-09-15.
\newblock \url{http://dlmf.nist.gov/}. F.~W.~J. Olver, A.~B. {Olde Daalhuis},
  D.~W. Lozier, B.~I. Schneider, R.~F. Boisvert, C.~W. Clark, B.~R. Miller,
  B.~V. Saunders, H.~S. Cohl, and M.~A. McClain, eds.

\bibitem{Jevicki:1998bm}
A.~Jevicki, M.~Mihailescu, and S.~Ramgoolam, ``{Gravity from CFT on S**N(X):
  Symmetries and interactions},''
  \href{http://dx.doi.org/10.1016/S0550-3213(00)00147-4}{{\em Nucl. Phys. B}
  {\bf 577} (2000)  47--72}, \href{http://arxiv.org/abs/hep-th/9907144}{{\tt
  arXiv:hep-th/9907144}}.

\bibitem{Schwimmer:1986mf}
A.~Schwimmer and N.~Seiberg, ``{Comments on the N=2, N=3, N=4 Superconformal
  Algebras in Two-Dimensions},''
\href{http://dx.doi.org/10.1016/0370-2693(87)90566-1}{{\em Phys. Lett.} {\bf
  B184} (1987)  191--196}.

\bibitem{SEN1990551}
A.~Sen, ``On the background independence of string field theory,''
  \href{http://dx.doi.org/https://doi.org/10.1016/0550-3213(90)90400-8}{{\em
  Nuclear Physics B} {\bf 345} (1990) no.~2, 551 -- 583}.
  \url{http://www.sciencedirect.com/science/article/pii/0550321390904008}.

\bibitem{Campbell:1990dz}
M.~Campbell, P.~C. Nelson, and E.~Wong, ``{Stress tensor perturbations in
  conformal field theory},''
  \href{http://dx.doi.org/10.1142/S0217751X9100232X}{{\em Int. J. Mod. Phys. A}
  {\bf 6} (1991)  4909--4924}.

\bibitem{Lima:2020urq}
A.~Lima, G.~Sotkov, and M.~Stanishkov, ``{Dynamics of R-neutral Ramond fields
  in the D1-D5 SCFT},'' \href{http://arxiv.org/abs/2012.08021}{{\tt
  arXiv:2012.08021 [hep-th]}}.

\end{thebibliography}\endgroup

\end{document}